\documentclass[lettersize,journal,twoside]{IEEEtran}
\usepackage{amsmath,amsfonts}
\usepackage{algorithmic}
\usepackage{algorithm}
\usepackage{array}
\usepackage[caption=false,font=footnotesize,labelfont=rm,textfont=rm]{subfig}
\usepackage{textcomp}
\usepackage{stfloats}
\usepackage{url}
\usepackage{verbatim}
\usepackage{graphicx}
\usepackage{cite}
\usepackage{booktabs,multirow,multicol,color}%
\begin{document}

\title{Community Detection for Heterogeneous Multiple Social Networks\\
}


\author{
Ziqing Zhu, 
Guan Yuan, 
Tao Zhou, 
and Jiuxin Cao
\thanks{This work was supported in part by the Fundamental Research Funds for the Central Universities under Grant 2023QN1082 (Corresponding
author: Ziqing Zhu and  Guan Yuan)}
\thanks{Ziqing Zhu is with the School of Computer Science and Technology,
China  University  of Mining and Technology, Xuzhou 221116, China (e-mail: zhuziqing@cumt.edu.cn)}
\thanks{Guan Yuan is with the School of Computer Science and Technology,
China University of Mining and Technology, Xuzhou 221116, China, 
also with the Jiangsu Key Laboratory of Mine Mechanical and Electrical
Equipment, China University of Mining and Technology, Xuzhou 221116, 
and also with the Digitization of Mine, Engineering Research Center, Ministry of Education, Xuzhou 221116, China (e-mail: yuanguan@cumt.edu.cn)}
\thanks{Tao Zhou is with the College of Computer and Information Engineering, Nanjing Tech University, Nanjing 211816, China (e-mail: t.zhou@njtech.edu.cn)}
\thanks{Jiuxin Cao is with the School of Cyber Science and Engineering, Southeast University, Nanjing 211189, China, 
and also with Purple Mountain Laboratories, Nanjing 211111, China (e-mail: jx.cao@seu.edu.cn)}
\thanks{------------------------}
}

\markboth{IEEE TRANSACTIONS ON COMPUTATIONAL SOCIAL SYSTEMS,~Vol.~XX, No.~X, XXX~20XX}
{Zhu \MakeLowercase{\textit{et al.}}: Community Detection for Heterogeneous Multiple Social Networks}
\IEEEpubid{0000--0000/00\$00.00~\copyright~2021 IEEE}

\maketitle

\begin{abstract}
The community plays a crucial role in understanding user behavior and network characteristics in social networks. 
Some users can use multiple social networks at once for a variety of objectives. These users are called overlapping users who bridge different social networks.
Detecting communities across multiple social networks is vital for interaction mining, information diffusion, and behavior migration analysis among networks. 
This paper presents a community detection method based on nonnegative matrix tri-factorization for multiple heterogeneous social networks, which formulates a common consensus matrix to represent the global fused community.
Specifically, the proposed method involves creating adjacency matrices based on network structure and content similarity, followed by alignment matrices which distinguish overlapping users in different social networks. 
With the generated alignment matrices, the method could enhance the fusion degree of the global community by detecting overlapping user communities across networks. 
The effectiveness of the proposed method is evaluated with new metrics on Twitter, Instagram, and Tumblr datasets. 
The results of the experiments demonstrate its superior performance in terms of community quality and community fusion.
\end{abstract}

\begin{IEEEkeywords}
community detection, social network,  matrix factorization, clustering, data mining
\end{IEEEkeywords}

\section{Introduction}
\IEEEPARstart{I}{n} recent years, online social networks have become more diverse and inclusive. 
They offer a wide range of content, from text to images, videos, and even location-based services. 
Social networks have also evolved to cater to individual requirements. 
Various social media platforms have emerged on the Internet, each with unique attributes. 
For example, Twitter and Weibo are similar to news media, while Facebook and WeChat focus more on daily life. 
Quora and Zhihu primarily emphasize knowledge sharing. 
Many users have accounts on different social media platforms to meet their personal needs. 
Such users are referred to as ``anchor users''\cite{zhang2014meta} or ``overlapping users''.
Understanding the structure of social networks relies heavily on integrating heterogeneous data scattered across various social networks. 
The heterogeneity of social networks brings a multi-dimensional research perspective. 
Overlapping users make it possible to carry out research for multiple social networks.

It is fascinating to see how people with similar interests and preferences gravitate toward each other on social media platforms. 
They frequently interact and show group structures through online social networks. 
These groups establish logical communities and contribute significantly to social networks. 
The study of the community structure in a social network can help in the analysis of the social network's structural characteristics\cite{cavallari2017learning,wang2017community,li2018community}, information diffusion pattern\cite{hu2015community,wang2019community}, etc. 
The research of community detection for multiple social networks has new challenges when compared with traditional community detection for a single social network, including: 
(1) How to obtain fused community across multiple social networks, which can still retain the intrinsic community characteristics from every single social network to the greatest extent possible. 
(2) How to design effective strategy to connect every social network by overlapping users, and enable each network to transfer its own community information which is used to reshape the community structure in other social networks for further facilitating community fusion.
(3) How to identify users across multiple social networks who exhibit similarities in behavior, preferences, and content in order to form a community, and choose different ways to treat overlapping and non-overlapping users, for example using different features or separate modeling, to highlight the distinctiveness of overlapping users.
\IEEEpubidadjcol

\begin{figure}[htb]
  \centerline{\includegraphics[width=0.32\textwidth]{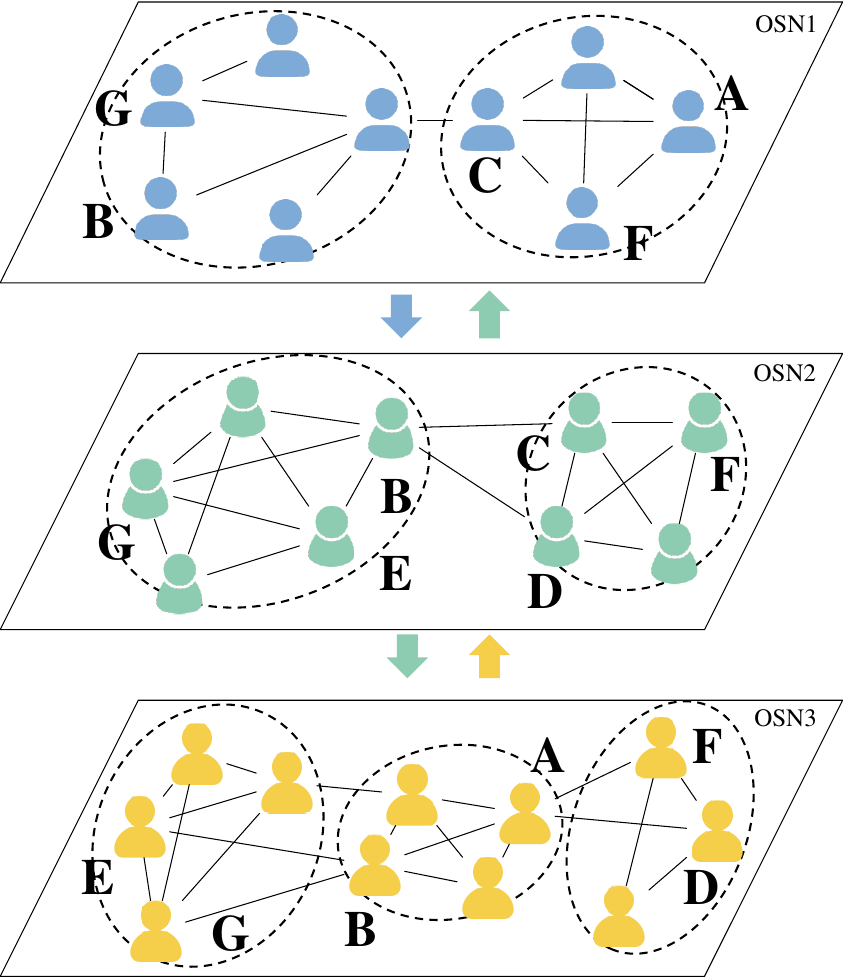}}
  \caption{Community structure across social networks. 
  In this figure, users A, B, C, D, E, F and G are overlapping users from the three social networks, OSN1, OSN2 and OSN3. 
  These users have different community structures in their social networks respectively. 
  The main difficulty in implementing community detection across multiple networks is integrating the communities of diverse networks with overlapping members as the core. 
  In OSN1, user C and user F are in the same community, 
  and in OSN3, user F and user D are in the same community. 
  Users C and D can be grouped in the same community by analyzing transitive relation. This division corresponds to the community division of OSN2's users C and D.
  However, in another case,  users A and B may be part of the same community in OSN3 but belong to different communities in OSN1. }
  
  \label{fig_multisocialcommunitystructure}
\end{figure}

Fig.\ref{fig_multisocialcommunitystructure} depicts the community structure of overlapping users in different social networks.
This figure illustrates a more generalized real-world scenario concerning multiple social networks.
It involves a group of overlapping users who possess accounts on multiple social networks. 
These overlapping users bridge the different social networks, forming \textbf{partially aligned} multiple social networks.
These social networks exhibit heterogeneity, where users can be divided into overlapping users and non-overlapping users.
This paper aims to create a fusion mechanism that can combine community structures from various social networks, 
extract and unify community structures across multiple social networks, 
and ensure that users with similar attributes are in the more consistent community.

\begin{figure}[htb]
  \centerline{\includegraphics[width=0.5\textwidth]{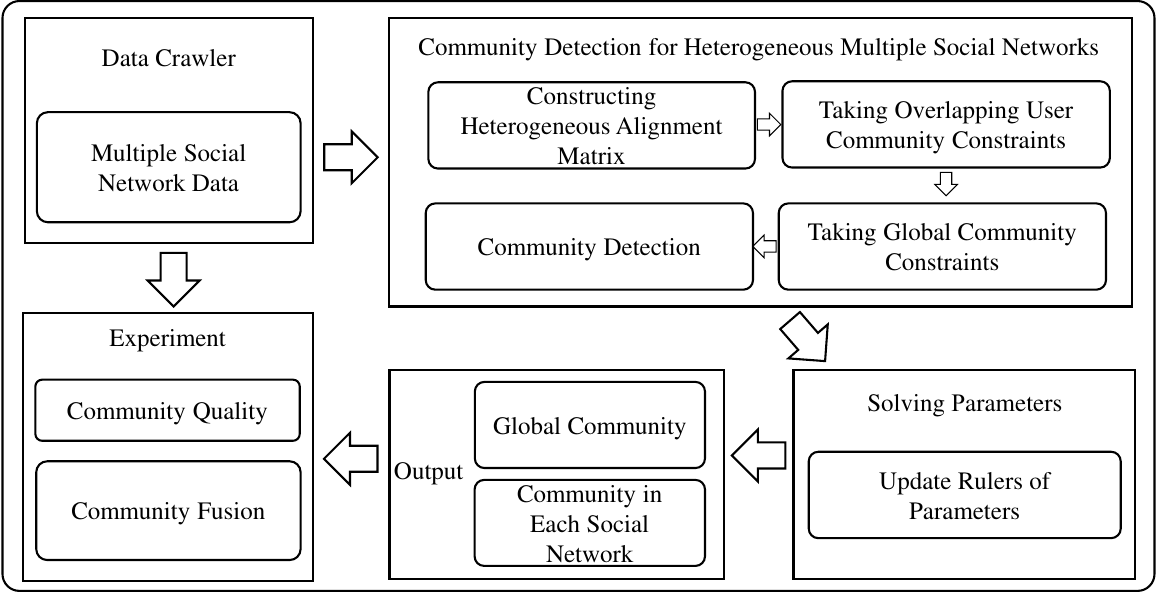}}
  \caption{Research framework}
  \label{fig_paperstructure}
\end{figure}

This paper proposes the Heterogeneous Multiple Social Networks Community Detection Model (HMCD). The research framework is shown in Fig.\ref{fig_paperstructure}. 
The main contributions of this paper are as follows.

1. In this paper, a community detection method based on nonnegative matrix tri-factorization is proposed for multiple social networks, which integrates the users' content and topology attributes and uses a common consensus matrix to obtain the fused community;

2. The community detection method allows for the identification of the global community structure across multiple social networks by introducing overlapping user community constraints;

3. This paper designs evaluation metrics for measuring community quality and fusion. It then conducts experiments on various social network datasets to demonstrate and analyze their effectiveness.

In the remainder of this paper, Section 2 reviews the related works; 
Section 3 introduces the research problem; 
Section 4 gives the community detection model; 
Section 5 describes the community detection algorithm; 
Section 6 evaluates the experiments; 
Finally, conclusions and future works are provided in Section 7.

\section{Related Works}
This section provides an overview of community detection works.
Related works of community for attributed networks and social networks are first reviewed.
Then, research works of community for multiplex networks and multiple social networks are described.
Lastly, this paper provides a summary and analysis of related works.
\subsection{Community Detection for Attributed Networks and Social Networks}
A community is commonly perceived as a group of closely related entities with similar attributes. The entities within the group are closely linked to each other.
However, their interaction with external objects is relatively weak\cite{chakraborty2017metrics}. 
Numerous studies have been undertaken to address the problems of community detection within homogeneous complex networks\cite{newman2003structure,girvan2002community,newman2004fast,yang2023lsha}.
Compared with traditional homogeneous complex networks, there is also a large amount of attribute data in addition to the topology information. 
Some existing methods not only exploit the topology structures of the network, but also take advantage of using the nodes' attribute information. 
Berahmand et al.\cite{berahmand2022graph} combined both the topological structure and attributed nodes of the network to discover community by constructing attribute augmented graph.
Berahmand et al.\cite{berahmand2024wsnmf} incorporated graph regularization and sparsity constraints to uphold the geometric structure of data points to obtain nodes'cluster in attributed networks.
He et al. \cite{he2024detecting} implement an autoencoder-style self-supervised learning model to detect community structure using multiple topics, which can adaptively fuse topology information and attribute information.
Jannesari al.\cite{jannesari2024novel} aligned structure and attribute spaces for clustering by combining clustering and community detection process. 
A distinctive feature of social networks is heterogeneity. Social networks have various multimedia elements.
Li et al. \cite{li2021community} proposed a method based on network representation learning, which combines local node embedding and global community embedding to achieve the goal of community detection.
Zarezadeh et al.\cite{zarezadeh2022dpnlp} have introduced a label propagation method based on the distances between nodes and their surrounding neighbors for community detection in social networks. 
In LBSN (Location-Based Social Network), 
Xu et al.\cite{xu2018community} considered the influence of user's friend relationships and time factors on users' access to specific locations and explored the location-based community structure of users. 
Kim et al.\cite{kim2020densely} have considered networks' online and offline characteristics in the community detection process. They emphasized the importance of considering both the online community relationships of users and the offline check-in density of users' locations within the community.
Ni et al. \cite{ni2023spatial} proposed a spatial aware local community detection algorithm based on dominance relation. They defined dominance relation inspired by multiobjective optimization.
In EBSN (Event-Based Social Network), 
Liu et al.\cite{liu2012event} considered the relationship between users' online and offline activities to study the community structure in the network. 
Li et al.\cite{li2020social} have integrated structure-based and behavior-based social influences to discover user communities. Structure-based social influence considers the online user topology, while behavior-based social influence considers the attributes of offline activities. 
\subsection{Community Detection for Multiplex Networks}
The intricate makeup of social networks, in terms of their diverse content and complex structure, renders them into multiplex networks.
A \textbf{single social network} can exhibit multi-view and multi-relational features at different levels, such as the user's following and forwarding networks. 
Ma et al.\cite{ma2018community} considered the connectivity between the layers of the multiplex social network. They designed a seed user selection mechanism at each layer, which was used as the core node to ensure the effectiveness of the final community detection algorithm. 
In addition, they proved the equivalence of kernel K-Means, spectral clustering, and non-negative matrix factorization-based community detection methods. 
Lv et al.\cite{lv2022community} proposed incorporating prior community information of nodes into their model and utilizing a semi-supervised joint symmetric non-negative matrix factorization algorithm to detect communities in multi-layer networks.

The community detection of multiplex networks has developed rapidly, but the main research object of these works is still essentially a single network. 
Some works have paid attention to the particularity of community detection across different networks. 
Luo et al.\cite{luo2020local} considered that different heterogeneous networks exhibit distinct community structures. They tried to design a local community detection model across networks based on random walks. 
Cao et al.\cite{cao2020mutual} analyzed the ``one-to-many'' relationship between news and tweets. Then they conducted topic community detection across the news web and Twitter. 
Ortiz-bouza et al.\cite{ortiz2022orthogonal} focused on the similarities and differences of communities across multiple networks. They have designed a method that enables the discovery of common communities across all networks and identifies private communities specific to each network.
These works mark overlapping entities across networks through entity similarity or explicit links. 
Based on these overlapping entities, they can realize the connection of different networks. 
\subsection{Community Detection for Multiple Social Networks}
In recent years, the functional differences between online social networks have made people use \textbf{multiple social networks}. 
Different social networks are naturally related by overlapping users. The behavioral consistency of people in various social networks produces similar data. 
As a result, the communities detected by each social network can be fused using these data. 
For the communities in multiple social networks, Nguyen et al.\cite{nguyen2015community} considered the differences in structural information among different social networks. They used the non-negative matrix factorization to detect social network community structure. 
Fan et al.\cite{fan2015overlapping} analyzed the connection edge characteristics of users in the community structure, found the community structure of overlapping users in every single network, and then aggregated them to obtain community across multiple social networks. 
Zhu et al.\cite{zhu2019community} designed a community detection method based on seed user diffusion, taking overlapping users as seed users for the characteristics of multiple social networks. 
Philip et al.\cite{philip2015mcd} studied the differences in overlapping users' community structures across multiple social networks. They made community detection based on the community confidence and community divergence characteristics of overlapping users.

\textbf{In summary}, community detection for multiple social networks and multi-layer networks is also related to multi-view clustering\cite{yao2019multi,luong2020novel,liu2013multi, he2014comment}. 
However, the latter focuses more on the attribute relationship between objects, not the topology. 
Multiple social networks have essentially the same concept compared with multiplex networks. 
The latter can be regarded as a particular case of the former. 
The users of the multiplex network can be viewed as wholly aligned. 
Overlapping users across multiple social networks are partially aligned. 
Because only some social network users can have accounts on different networks in the real world. 
From a macro perspective, communities in multi-layer networks focus more on community complementation between different layers\cite{magnani2021community}. 
The community for multiple social networks is based on the assumption of consistency. 
Existing research fails to comprehensively capture the fused community from a global perspective, and there are still few works on community detection for multiple social networks.
As a result, this paper will detect the global community by adjusting the similarity of overlapping users' community structures in each social network. 
It is possible to strengthen the fusion performance of communities and obtain a global community structure.

\section{Problem formulation}
This section first gives the definitions of related concepts and, on this basis, formally describes the problems studied in this paper.

We will use a matrix form to represent user relationships and community structure. 
The related concepts are defined as follows.

\textbf{Definition 1. Multiple Social Networks.}
This paper defines the social network set $\boldsymbol{{\rm{S}}}$, and user set $\boldsymbol{{\rm{U}}}$. 
$\boldsymbol{{\rm{U}}}^s$ represents the set of users in the social network $s(s \in \boldsymbol{{\rm{S}}})$ and $\boldsymbol{{\rm{U}}}_o^s$ represents the set of overlapping users in the social network $s$.

\textbf{Definition 2. Adjacency Matrix.}  
Network topology attributes $t$ and content attributes $e$ are important aspects of social networks. 
The user adjacency matrix for attribute $k$ is represented as $\boldsymbol{{\rm{M}}}_k^s$, and the overlapping user adjacency matrix for attribute $k$ is represented as $\boldsymbol{{\rm{M}}}_{k,o}^s$.

\textbf{Definition 3. Community Structure Matrix.}
The community structure matrix $\boldsymbol{{\rm{X}}}^s_k$ represents the community structure created by users with attribute $k$ in social network $s$. 
Similarly, the community structure matrix $\boldsymbol{{\rm{X}}}^s_{k,o}$ represents the community structure created by overlapping users with attribute $k$ in social network $s$. 
Each element in the matrix's row represents the user's degree of membership in the community.
We determine the community to which a user $m$ belongs by finding the index $l$ that maximizes $(\boldsymbol{{\rm{X}}})_{m,l}$. 
The global community structure matrix in this paper is denoted as $\boldsymbol{{\rm{C}}}$. 

We have a dataset of multiple social networks $\boldsymbol{{\rm{S}}}$ and the user and overlapping user adjacency matrices $\boldsymbol{{\rm{M}}}_k^s$, $\boldsymbol{{\rm{M}}}_{k,o}^s$ for the social network attributes $k$. 
The objective of this paper is to investigate the acquisition of community structure for multiple social networks, enabling a global description of users' community structures. 
The Tab.\ref{tab_Descriptions of symbol_com} provides a summary of the symbols used in this paper.

\begin{table}[htb]
  \caption{Descriptions of Symbol}
  \newcommand{\tabincell}[2]{\begin{tabular}{@{}#1@{}}#2\end{tabular}}
  \begin{center}
  \resizebox{0.43\textwidth}{!}{%
  \begin{tabular}{c|c}
  \toprule
  \textbf{Symbol} & \textbf{Description}\\
  \midrule
  $\boldsymbol{{\rm{S}}}$, $\boldsymbol {{\rm{U}}}$& \tabincell{c}{social network set, user set}\\
  \hline
  $\boldsymbol{{\rm{U}}}^s$, $\boldsymbol{{\rm{U}}}^s_o$& \tabincell{c}{user set in social network $s$, \\overlapping user set in social network $s$}\\
  \hline
  $\boldsymbol{{\rm{M}}}^s_k$, $\boldsymbol{{\rm{M}}}^s_{k,o}$&\tabincell{c}{user adjacency matrix,\\overlapping user adjacency matrix}\\
  \hline
  $\boldsymbol{{\rm{X}}}^s_k$, $\boldsymbol{{\rm{X}}}^s_{k,o}$&\tabincell{c}{user community structure matrix, \\overlapping user community structure matrix}\\
  \hline
  $\boldsymbol{{\rm{D}}}^s_k$, $\boldsymbol{{\rm{D}}}^s_{k,o}$ &community-community relation matrix\\
  \hline
  $\boldsymbol{{\rm{C}}}$&global community structure matrix\\
  \hline
  $\boldsymbol{{\rm{H}}}^s$,$\boldsymbol{{\rm{T}}}^s$&alignment matrix\\
  \hline
  $\boldsymbol{{\rm{Q}}}_{k}^s$, $\boldsymbol{{\rm{Q}}}_{k,o}^s$&normalization matrix\\
  \hline
  $K$&the number of communities \\
  \hline
  $k$&social network attributes(topology, content)\\
  \hline
  $\alpha_k^s$, $\beta_k^s$&term weight of social networks\\
  \hline
  $\gamma_k^s$ &term weight of overlapping user community\\
  \hline
  $\theta_k^s$ &term weight of global community\\
  \bottomrule
  \end{tabular}}
  \label{tab_Descriptions of symbol_com}
  \end{center}
\end{table}

\section{Global Community Detection Model}
\begin{figure*}
  \centerline{\includegraphics[width=0.88\textwidth]{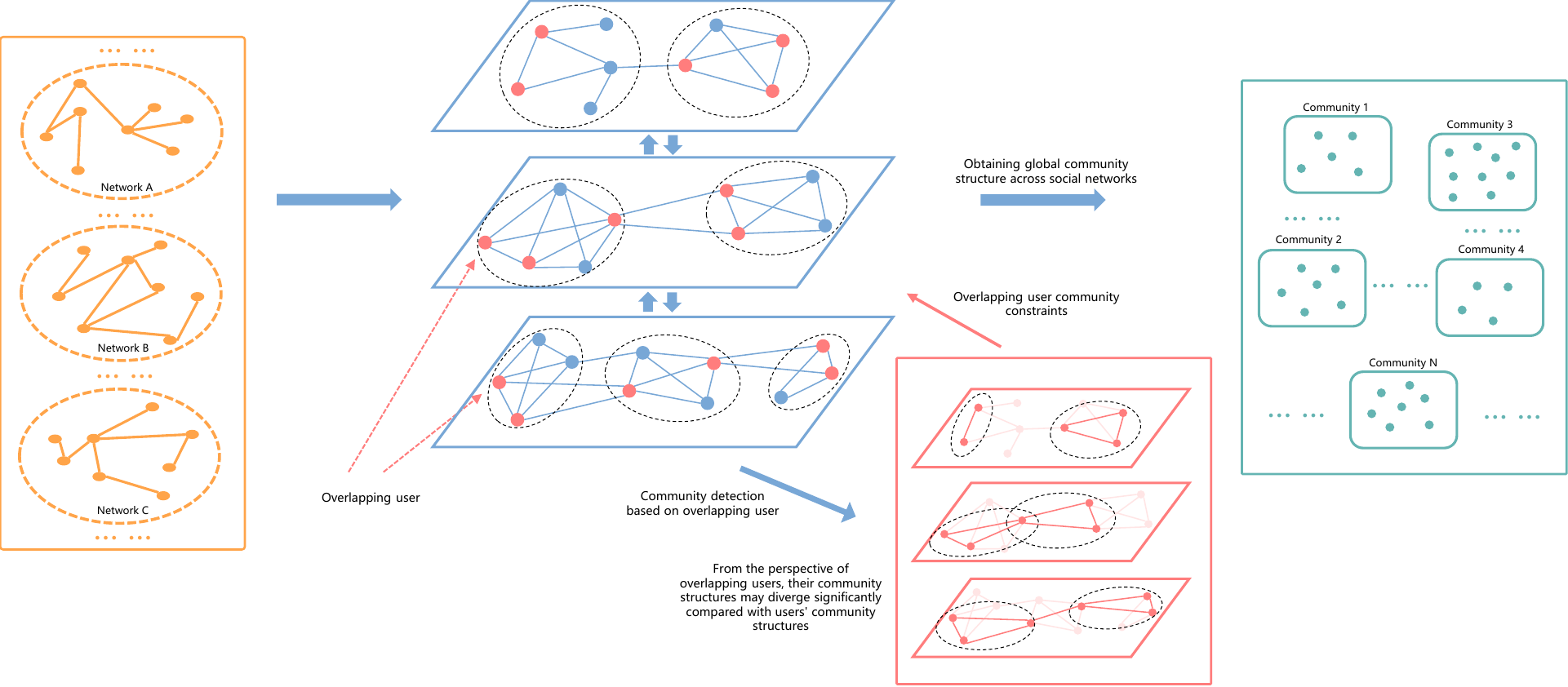}}
  \caption{Illustration of the proposed community detection model across multiple social networks. 
  Our approach primarily takes into account some overlapping users across multiple social networks. 
  Therefore these networks are partially aligned. 
  We posit that the cohort of overlapping users forms the essence of community across these networks, with the community forged among these users potentially exhibiting significant disparities from those among regular users. 
  Furthermore, the community of overlapping users serves as a pivotal component for fostering community fusion.
  \label{fig_illustrationmodel}}  
\end{figure*}

A community detection model with a global fusion structure will be developed in this paper. 
We associate the community structure of different social networks with overlapping users. 
Our main idea is described as Fig.\ref{fig_illustrationmodel}.
\subsection{Modeling Attribute Relation Between Users}
\subsubsection{Adjacency Matrix Factorization Based on Topology Attributes}
The topology attributes in social networks are usually divided into two types: 
The first type is one-way following topology in platforms, such as Weibo and Twitter, and the corresponding network topology is a directed graph. 
The second type is bidirectional following topology in platforms, such as WeChat and Facebook. 
The corresponding network topology is an undirected graph.
Specifically, the network topology is built based on the users' following relations, and then the user-user adjacency matrix is obtained.

Taking the asymmetric matrix constructed by a directed graph as an example, its tri-factorization form\cite{wang2011community} of the community structure is shown in Eq.\ref{eq_topologydecomposition}:
\begin{equation}\label{eq_topologydecomposition}
  \min_{\boldsymbol{{\rm{X}}}_t^s} \Vert \boldsymbol{{\rm{M}}}_t^s -  \boldsymbol{{\rm{X}}}_t^s  \boldsymbol{{\rm{D}}}_t^s {\boldsymbol{{\rm{X}}}_t^s}^{\rm{T}} \Vert_F^2 \quad s.t.\boldsymbol{{\rm{X}}}_t^s \geq 0, \boldsymbol{{\rm{D}}}_t^s \geq 0
\end{equation}
where $\boldsymbol{{\rm{X}}}_t^s$ represents the community structure matrix of users based on topology attributes $t$ in social network$s$.
$\boldsymbol{{\rm{D}}}_t^s$ represents the community-community relation matrix.
\subsubsection{Adjacency Matrix Factorization based on Content Attributes}
In social networks, users have similar content interests and preferences in the same community. 
This paper calculates the semantic similarity of users' content. 
Then the user-user adjacency matrix can be obtained, which is a symmetric matrix, and its factorization form\cite{wang2011community} about the community structure is shown in Eq.\ref{eq_contentdecomposition}:

\begin{equation}\label{eq_contentdecomposition}
  \min_{\boldsymbol{{\rm{X}}}_e^s} \Vert \boldsymbol{{\rm{M}}}_e^s -  \boldsymbol{{\rm{X}}}_e^s  {\boldsymbol{{\rm{X}}}_e^s}^{\rm{T}} \Vert_F^2 \quad s.t.\boldsymbol{{\rm{X}}}_e^s \geq 0
\end{equation}
where $\boldsymbol{{\rm{X}}}_e^s$ represents the community structure matrix of the user under the content attribute $e$ of the social network $s$.

The factorization of a symmetric matrix can also be transformed into the factorization of an asymmetric matrix.
Furthermore, the community-community relation matrix, which provides more degrees of freedom, can increase the accuracy of the factorization approximation when the community structure matrix provides community membership information.
The following sections will uniformly use the matrix tri-factorization of the asymmetric form to describe the community detection model.

\subsection{Overlapping User Community}
Users who belong to multiple social networks are considered as overlapping users and serve as bridges connecting those networks. The relationships and connections between these overlapping users are crucial in merging the community structures of networks.
Therefore, this paper designs an alignment matrix to mark overlapping users in social networks to describe their community structure.

\begin{figure}[htb]
  \centerline{\includegraphics[width=0.33\textwidth]{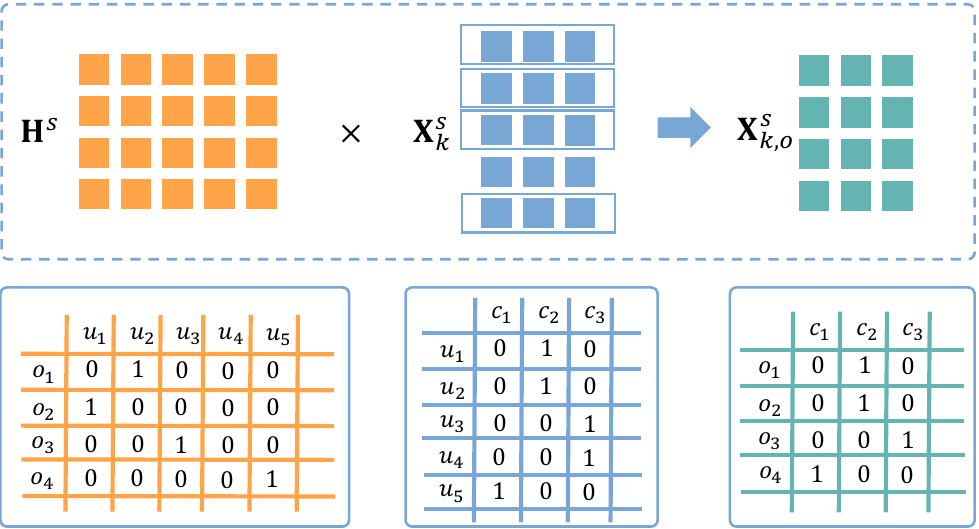}}
  \caption{In alignment matrix operations, the value of an element in the matrix indicates whether the row user $u_i$ and the column user $o_j$ are the same users.}
  \label{fig_matrixmanipulateion}
\end{figure}

Specifically, on the premise of knowing the overlapping user identities, this paper constructs an alignment matrix $\boldsymbol{{\rm{H}}}^s$, as shown in Fig.\ref{fig_matrixmanipulateion}, whose columns represent users in the social network $s$, and the rows are identified as overlapping users in the social network $s$. 
The element value in the matrix indicates whether the row user $u_i$ and the column user $o_j$ are the same users. If they are the same user, the value is set to 1; otherwise, it is 0.
In this way, the user community structure matrix $\boldsymbol{{\rm{X}}}_k^s$ of the social network $s$ can be converted into an overlapping user community structure matrix $\boldsymbol{{\rm{X}}}_{k,o}^s$ through the alignment matrix $\boldsymbol{{\rm{H}}}^s$, that is $\boldsymbol{{\rm{X}}}_{k,o}^s=\boldsymbol{{\rm{H}}}^s \boldsymbol{{\rm{X}}}_k^s$.

According to the attribute $k$ of the social network $s$, the adjacency matrix of users and overlapping users are constructed respectively. 
Through the alignment of users, the community detection for social networks $s$ can be obtained by the Eq.\ref{eq_overlappingusercommunitydecomposition}.

\begin{equation}\label{eq_overlappingusercommunitydecomposition}
  \begin{aligned}
    \min_{\boldsymbol{{\rm{X}}}_k^s}\hspace*{0.3em} &\alpha^s_k \Vert \boldsymbol{{\rm{X}}}_k^s -  \boldsymbol{{\rm{X}}}_k^s  \boldsymbol{{\rm{D}}}_k^s  {\boldsymbol{{\rm{X}}}_k^s}^{\rm{T}} \Vert_F^2  +
     \beta^s_k \Vert \boldsymbol{{\rm{X}}}_{k,o}^s -  \boldsymbol{{\rm{X}}}_{k,o}^s  \boldsymbol{{\rm{D}}}_{k,o}^s  {\boldsymbol{{\rm{X}}}_{k,o}^s}^{\rm{T}} \Vert_F^2 \\+
    & \gamma^s_k \Vert \boldsymbol{{\rm{H}}}^s\boldsymbol{{\rm{X}}}_k^s-\boldsymbol{{\rm{X}}}_{k,o}^s \Vert_F^2,\quad s.t.\boldsymbol{{\rm{X}}}_k^s \geq 0,\boldsymbol{{\rm{X}}}_{k,o}^s \geq 0
  \end{aligned}
\end{equation}
where, $\boldsymbol{{\rm{X}}}_k^s$ and $\boldsymbol{{\rm{X}}}_{k,o}^s$ represent the above two community structures of the social network $s$ respectively. 
$\boldsymbol{{\rm{X}}}_k^s$ and $\boldsymbol{{\rm{X}}}_{k,o}^s$ represent the community-community relation matrix respectively. 
The balance factor $\alpha^s_k$, $\beta^s_k$, and $\gamma^s_k$ are used to adjust the weight of the overlapping user community structure so that the modeling can reflect the importance of overlapping users in multiple social networks.
$\Vert \boldsymbol{{\rm{H}}}^s\boldsymbol{{\rm{X}}}_k^s-\boldsymbol{{\rm{X}}}_{k,o}^s \Vert_F^2$ aims to minimize the differences in community structures between users and overlapping users to facilitate the fusion of community detection.

\subsection{Community Structure Across Social Networks}
Our goal is to create a global community by combining different social networks. 
To achieve this, we will include global community constraints in the optimization objective for community detection, and provide a unified description of all community members from a global perspective.

Similarly, we construct an alignment matrix of users $\boldsymbol{{\rm{T}}}^s$ to mark the corresponding users in the social network $s$ for the global community. 
Its rows are users in the social network $s$, and columns are users in the global community. 

Every social network has different heterogeneous information, and this will affect the fusion of community structure matrix $\boldsymbol{{\rm{X}}}_k^s$ across social networks due to the difference in the value dimension or range of the attribute data. 
Referring to the work\cite{liu2013multi}, this paper normalizes the user community structure matrix $\boldsymbol{{\rm{X}}}_k^s$.
Specifically, let non-negative matrices follow $ \boldsymbol{{\rm{M}}}=\boldsymbol{{\rm{E}}} \boldsymbol{{\rm{F}}}^{\rm{T}} $, if $\sum _i \boldsymbol{{\rm{E}}}_{i,.} =1 $, since $\boldsymbol{{\rm{M}}}_{.,j} = \sum _m  \boldsymbol{{\rm{E}}}_{.,m} \boldsymbol{{\rm{F}}}_{j,m} $ , we can get 
$\Vert \boldsymbol{{\rm{M}}} \Vert_1 =\Vert \sum _j \boldsymbol{{\rm{M}}}_{.,j}  \Vert_1
=  \Vert  \sum _m  \boldsymbol{{\rm{E}}}_{.,m}  \sum _j \boldsymbol{{\rm{F}}}_{j,m}  \Vert_1 
= \Vert  \sum _m  \sum _j \boldsymbol{{\rm{F}}}_{j,m}  \Vert_1
= \Vert \boldsymbol{{\rm{F}}} \Vert_1 $.

Therefore, the user adjacency matrix $\Vert \boldsymbol{{\rm{M}}}_k^s \Vert$ is first normalized so that the sum of all elements in the matrix is 1, that is $\Vert \boldsymbol{{\rm{M}}}_k^s \Vert_1=1$. 
Then, we construct the normalized matrix $\boldsymbol{{\rm{Q}}}_k^s$.
\begin{equation}\label{eq_qfunction}
  \boldsymbol{{\rm{Q}}}_k^s=Diag(\sum_i (\boldsymbol{{\rm{X}}}_k^s\boldsymbol{{\rm{D}}}_k^s)_{i,1}\ldots \sum_i (\boldsymbol{{\rm{X}}}_k^s\boldsymbol{{\rm{D}}}_k^s)_{i,N})
\end{equation}
where $Diag()$ denotes a diagonal matrix. $\boldsymbol{{\rm{Q}}}_k^s$ can make $\sum_i (\boldsymbol{{\rm{X}}}_k^s\boldsymbol{{\rm{D}}}_k^s(\boldsymbol{{\rm{Q}}}_k^s)^{-1})_{i,.}=1$

In the model training process, this paper transforms the parameters of the model through $(\boldsymbol{{\rm{X}}}_k^s)^{new}\leftarrow \boldsymbol{{\rm{X}}}_k^s\boldsymbol{{\rm{Q}}}_k^s$ and $(\boldsymbol{{\rm{D}}}_k^s)^{new}  \leftarrow  (\boldsymbol{{\rm{Q}}}_k^s)^{-1} \boldsymbol{{\rm{D}}}_k^s (\boldsymbol{{\rm{Q}}}_k^s)^{-1}$ to make $\Vert (\boldsymbol{{\rm{X}}}_k^s)^{new} \Vert_1=  \Vert \boldsymbol{{\rm{M}}}_k^s \Vert_1 =1$.
Furthermore, this paper obtains the global community constraint:
\begin{equation}\label{eq_difffunction}
  D(\boldsymbol{{\rm{X}}}^s_k,\boldsymbol{{\rm{C}}})=\sum_{\boldsymbol{{\rm{S}}}}\sum_{\boldsymbol{{\rm{K}}}} \Vert \boldsymbol{{\rm{X}}}_k^s\boldsymbol{{\rm{Q}}}_k^s-\boldsymbol{{\rm{T}}}^s\boldsymbol{{\rm{C}}} \Vert^2_F
\end{equation}

This equation describes approximation between the global community and intrinsic community in each social network. The model can obtain global community $\boldsymbol{{\rm{C}}}$, which represents fused community structure on all social networks.
For the overlapping user community structure $\boldsymbol{{\rm{M}}}_{k,o}^s$, normalization can also be carried out as $\Vert \boldsymbol{{\rm{M}}}_{k,o}^s \Vert_1=1$.
We multiply the community structure matrix $\boldsymbol{{\rm{X}}}_{k,o}^s$ and the normalization matrix $\boldsymbol{{\rm{Q}}}_{k,o}^s$.
\begin{equation}\label{eq_qofunction}
  \boldsymbol{{\rm{Q}}}_{k,o}^s=Diag(\sum_i (\boldsymbol{{\rm{X}}}_{k,o}^s\boldsymbol{{\rm{D}}}_{k,o}^s)_{i,1}\ldots \sum_i (\boldsymbol{{\rm{X}}}_{k,o}^s\boldsymbol{{\rm{D}}}_{k,o}^s)_{i,N})
\end{equation}

Finally, this paper can get the objective optimization function of the model:
\begin{equation}\label{eq_globalcommunitydecomposition}
  \begin{aligned}
    \min_{\boldsymbol{{\rm{X}}}_k^s,C} \sum_{\boldsymbol{{\rm{S}}}} \sum_{\boldsymbol{{\rm{K}}}}& \alpha^s_k \Vert \boldsymbol{{\rm{M}}}_k^s -  \boldsymbol{{\rm{X}}}_k^s  \boldsymbol{{\rm{D}}}_k^s  {\boldsymbol{{\rm{X}}}_k^s}^{\rm{T}} \Vert_F^2\\ + 
    &\beta^s_k \Vert \boldsymbol{{\rm{M}}}_{k,o}^s -  \boldsymbol{{\rm{X}}}_{k,o}^s  \boldsymbol{{\rm{D}}}_{k,o}^s  {\boldsymbol{{\rm{X}}}_{k,o}^s}^{\rm{T}} \Vert_F^2 \\+
    & \gamma^s_k \Vert \boldsymbol{{\rm{H}}}^s\boldsymbol{{\rm{X}}}_k^s\boldsymbol{{\rm{Q}}}_k^s-\boldsymbol{{\rm{X}}}_{k,o}^s\boldsymbol{{\rm{Q}}}_{k,o}^s \Vert_F^2 \\+ &\theta_k^s \Vert \boldsymbol{{\rm{X}}}_k^s\boldsymbol{{\rm{Q}}}_k^s-\boldsymbol{{\rm{T}}}^s\boldsymbol{{\rm{C}}} \Vert^2_F \\
    \quad s.t.\boldsymbol{{\rm{X}}}_k^s \geq 0,&\boldsymbol{{\rm{X}}}_{k,o}^s \geq 0, \boldsymbol{{\rm{C}}} \geq 0,  \boldsymbol{{\rm{D}}}_{k,o}^s \geq 0,\boldsymbol{{\rm{D}}}_{k,o}^s \geq 0
  \end{aligned}
\end{equation}

where, $\boldsymbol{{\rm{M}}}_k^s$ and $\boldsymbol{{\rm{M}}}_{k,o}^s$ are the adjacency matrices of all users and overlapping users under the attribute $k$ of the social network $s$ respectively. 
$\boldsymbol{{\rm{D}}}_k^s$ and $\boldsymbol{{\rm{D}}}_{k,o}^s$ represent the community-community relationship matrices. 
$\boldsymbol{{\rm{Q}}}_k^s$ and $\boldsymbol{{\rm{Q}}}_{k,o}^s$ are normalized matrices. 
$\boldsymbol{{\rm{C}}}$ is the user global community structure matrix.

\section{Community Detection Algorithm for Multiple Social Networks}
This section gives the solution method of the model parameters and the community detection algorithm based on the community detection model.

\subsection{model parameter solution}
The optimization objective of Eq.\ref{eq_lagrangeglobalcommunitydecomposition} can be solved by constructing the Lagrangian function. 
The Lagrangian function is shown as follows.
\begin{equation}\label{eq_lagrangeglobalcommunitydecomposition}
  \begin{aligned}
     L( \boldsymbol{{\rm{X}}},  \boldsymbol{{\rm{D}}},\boldsymbol{{\rm{C}}}) &=
    \sum_{\boldsymbol{{\rm{S}}}} \sum_{\boldsymbol{{\rm{K}}}} \alpha^s_k \Vert \boldsymbol{{\rm{M}}}_k^s -  \boldsymbol{{\rm{X}}}_k^s  \boldsymbol{{\rm{D}}}_k^s  {\boldsymbol{{\rm{X}}}_k^s}^{\rm{T}} \Vert_F^2\\
    & + \beta^s_k \Vert \boldsymbol{{\rm{M}}}_{k,o}^s -  \boldsymbol{{\rm{X}}}_{k,o}^s  \boldsymbol{{\rm{D}}}_{k,o}^s  {\boldsymbol{{\rm{X}}}_{k,o}^s}^{\rm{T}} \Vert_F^2 \\
    & + \gamma^s_k \Vert \boldsymbol{{\rm{H}}}^s\boldsymbol{{\rm{X}}}_k^s\boldsymbol{{\rm{Q}}}_k^s - \boldsymbol{{\rm{X}}}_{k,o}^s\boldsymbol{{\rm{Q}}}_{k,o}^s \Vert_F^2 \\
    & + \theta_k^s \Vert \boldsymbol{{\rm{X}}}_k^s\boldsymbol{{\rm{Q}}}_k^s-\boldsymbol{{\rm{T}}}^s\boldsymbol{{\rm{C}}} \Vert^2_F 
     -  tr(\boldsymbol{{\rm{\mu }}}{\boldsymbol{{\rm{X}}}_k^s}^{\rm{T}})\\
    & -tr(\boldsymbol{{\rm{\rho }}}{\boldsymbol{{\rm{X}}}_{k,o}^s} ^{\rm{T}})-tr(\boldsymbol{{\rm{\pi }}}\boldsymbol{{\rm{C}}} ^{\rm{T}})
     -tr(\boldsymbol{{\rm{\omega}}}{\boldsymbol{{\rm{D}}}_k^s} ^{\rm{T}})\\
    & -tr(\boldsymbol{{\rm{\varphi}}}{\boldsymbol{{\rm{D}}}_{k,o}^s}^{\rm{T}})
  \end{aligned}
\end{equation}
where $tr()$ represents the trace of the matrix.

To solve the optimization objective of Eq.\ref{eq_globalcommunitydecomposition}, it can be equivalent to making the Lagrangian function satisfy the following conditions:
\begin{equation}\label{eq_solvercondition}
  \left\{
  \begin{aligned}
   &\frac{\partial L}{\boldsymbol{{\rm{X}}}_k^s}=0, \frac{\partial L}{\boldsymbol{{\rm{X}}}_{k,o}^s}=0, \frac{\partial L}{\boldsymbol{{\rm{D}}}_k^s}=0, \frac{\partial L}{\boldsymbol{{\rm{D}}}_{k,o}^s}=0,\frac{\partial L}{\boldsymbol{{\rm{C}}}}=0\\
   &\mbox{KKT: }(\boldsymbol{{\rm{\mu }}})_{ij}(\boldsymbol{{\rm{X}}}_k^s)_{ij}=0,(\boldsymbol{{\rm{\rho }}})_{ij}(\boldsymbol{{\rm{X}}}_{k,o}^s)_{ij}=0,\\
   &(\boldsymbol{{\rm{\pi }}})_{ij}(\boldsymbol{{\rm{C}}})_{ij}=0,(\boldsymbol{{\rm{\omega}}})_{ij}({\boldsymbol{{\rm{D}}}_k^s})_{ij}=0,(\boldsymbol{{\rm{\varphi}}})_{ij}(\boldsymbol{{\rm{D}}}_{k,o}^s)_{ij}=0
  \end{aligned}
  \right.
\end{equation}
 
Therefore, on this basis, this paper can solve $\boldsymbol{{\rm{X}}}_k^s$, $\boldsymbol{{\rm{X}}}_{k,o}^s$, $\boldsymbol{{\rm{D}}}_k^s$, $\boldsymbol{{\rm{D}}}_{k,o}^s$, $\boldsymbol{{\rm{C}}}$.

1) For $\boldsymbol{{\rm{X}}}_k^s$, this paper first performs the normalization operation $\boldsymbol{{\rm{X}}}_k^s\leftarrow \boldsymbol{{\rm{X}}}_k^s\boldsymbol{{\rm{Q}}}_k^s$, $\boldsymbol{{\rm{D}}}_k^s\leftarrow (\boldsymbol{{\rm{Q}}}_k^s)^{-1}\boldsymbol{{\rm{D}}}_k^s(\boldsymbol{{\rm{Q}}}_k^s)^{-1}$.
Then the value of the term $\Vert \boldsymbol{{\rm{M}}}_k^s -  \boldsymbol{{\rm{X}}}_k^s  \boldsymbol{{\rm{D}}}_k^s  {\boldsymbol{{\rm{X}}}_k^s}^{\rm{T}} \Vert_F^2$ in Eq.\ref{eq_lagrangeglobalcommunitydecomposition} will not change, 
and $\boldsymbol{{\rm{X}}}_k^s\boldsymbol{{\rm{Q}}}_k^s$ can be replaced by $\boldsymbol{{\rm{X}}}_k^s$. 
Considering the KKT condition, we can obtain the update rule of parameter iteration\cite{wang2011community,seung2001algorithms}:
\begin{equation}\label{eq_xsolver}
  \begin{aligned}
   &\quad \quad \quad \quad (\boldsymbol{{\rm{X}}}_k^s)_{m,n}=(\boldsymbol{{\rm{X}}}_k^s)_{m,n}\sqrt[4]{\frac{f(\boldsymbol{{\rm{X}}}_k^s)_0}{f(\boldsymbol{{\rm{X}}}_k^s)_1} } \\
   &\enspace\enspace \enspace  f(\boldsymbol{{\rm{X}}}_k^s)_0 = (\alpha^s_k({\boldsymbol{{\rm{M}}}_k^s}^{\rm{T}}\boldsymbol{{\rm{X}}}_k^s\boldsymbol{{\rm{D}}}_k^s+\boldsymbol{{\rm{M}}}_k^s\boldsymbol{{\rm{X}}}_k^s{\boldsymbol{{\rm{D}}}_k^s}^{\rm{T}})\\
   &\quad \quad \quad \quad \quad +\gamma^s_k{\boldsymbol{{\rm{H}}}^s}^{\rm{T}}\boldsymbol{{\rm{X}}}_{k,o}^s\boldsymbol{{\rm{Q}}}_{k,o}^s+\theta^s_k\boldsymbol{{\rm{T}}}^s\boldsymbol{{\rm{C}}})_{m,n} \\
   &\enspace \enspace \enspace  f(\boldsymbol{{\rm{X}}}_k^s)_1 = (\alpha^s_k(\boldsymbol{{\rm{X}}}_k^s{\boldsymbol{{\rm{D}}}_k^s}^{\rm{T}}{\boldsymbol{{\rm{X}}}_k^s}^{\rm{T}}\boldsymbol{{\rm{X}}}_k^s\boldsymbol{{\rm{D}}}_k^s + \boldsymbol{{\rm{X}}}_k^s\boldsymbol{{\rm{D}}}_k^s{\boldsymbol{{\rm{X}}}_k^s}^{\rm{T}}\boldsymbol{{\rm{X}}}_k^s{\boldsymbol{{\rm{D}}}_k^s}^{\rm{T}})\\
   &\quad  \quad \quad \quad \quad +\gamma^s_k{\boldsymbol{{\rm{H}}}^s}^{\rm{T}}(\boldsymbol{{\rm{H}}}^s\boldsymbol{{\rm{X}}}_k^s)+\theta_k^s\boldsymbol{{\rm{X}}}_k^s)_{m,n}   
  \end{aligned}
\end{equation} 

2) For $\boldsymbol{{\rm{X}}}_{k,o}^s$, this paper also normalizes $\boldsymbol{{\rm{X}}}_{k,o}^s \leftarrow \boldsymbol{{\rm{X}}}_{k,o}^s\boldsymbol{{\rm{Q}}}_{k,o}^s$, and obtains the iterative update rule through the Lagrangian function of $\boldsymbol{{\rm{X}}}_{k,o}^s$ as described above:
\begin{equation}\label{eq_xosolver}
  \begin{aligned}
  &\quad \quad \quad \quad (\boldsymbol{{\rm{X}}}_{k,o}^s)_{m,n}=(\boldsymbol{{\rm{X}}}_{k,o}^s)_{m,n}\sqrt[4]{\frac{f(\boldsymbol{{\rm{X}}}_{k,o}^s)_0}{f(\boldsymbol{{\rm{X}}}_{k,o}^s)_1} } \\
  &\quad   f(\boldsymbol{{\rm{X}}}_{k,o}^s)_0 = (\beta^s_k({\boldsymbol{{\rm{M}}}_{k,o}^s}^{\rm{T}}\boldsymbol{{\rm{X}}}_{k,o}^s\boldsymbol{{\rm{D}}}_{k,o}^s + \boldsymbol{{\rm{M}}}_{k,o}^s\boldsymbol{{\rm{X}}}_{k,o}^s{\boldsymbol{{\rm{D}}}_{k,o}^s}^{\rm{T}})\\
  &\quad \quad \quad \quad \quad    +\gamma^s_k{\boldsymbol{{\rm{H}}}^s}\boldsymbol{{\rm{X}}}_k^s\boldsymbol{{\rm{Q}}}_k^s)_{m,n} \\
  &\quad   f(\boldsymbol{{\rm{X}}}_{k,o}^s)_1 = (\beta^s_k(\boldsymbol{{\rm{X}}}_{k,o}^s{\boldsymbol{{\rm{D}}}_{k,o}^s}^{\rm{T}}{\boldsymbol{{\rm{X}}}_{k,o}^s}^{\rm{T}}\boldsymbol{{\rm{X}}}_{k,o}^s\boldsymbol{{\rm{D}}}_{k,o}^s+\\
  &\quad \quad \quad \quad \quad \quad \quad  \boldsymbol{{\rm{X}}}_{k,o}^s\boldsymbol{{\rm{D}}}_{k,o}^s{\boldsymbol{{\rm{X}}}_{k,o}^s}^{\rm{T}}\boldsymbol{{\rm{X}}}_{k,o}^s{\boldsymbol{{\rm{D}}}_{k,o}^s}^{\rm{T}})  +\gamma^s_k\boldsymbol{{\rm{X}}}_{k,o}^s)_{m,n}
  \end{aligned}
\end{equation} 

3) For $\boldsymbol{{\rm{D}}}_k^s$, we also need to pay attention to the $\boldsymbol{{\rm{D}}}_k^s$ variable of $\boldsymbol{{\rm{Q}}}_k^s$ in Eq.\ref{eq_globalcommunitydecomposition}. By utilizing the Lagrangian function of $\boldsymbol{{\rm{D}}}_k^s$, the iterative update form can be obtained:
\begin{equation}\label{eq_dsolver}
  \begin{aligned}
  & \quad \quad \quad \quad (\boldsymbol{{\rm{D}}}_k^s)_{m,n}=(\boldsymbol{{\rm{D}}}_k^s)_{m,n}\sqrt{\frac{f(\boldsymbol{{\rm{D}}}_k^s)_0}{f(\boldsymbol{{\rm{D}}}_k^s)_1} } \\
  & \quad f(\boldsymbol{{\rm{D}}}_k^s)_0 = \alpha^s_k({\boldsymbol{{\rm{X}}}_k^s}^{\rm{T}}\boldsymbol{{\rm{M}}}_k^s\boldsymbol{{\rm{X}}}_k^s)_{m,n}\\
  &\quad \quad  \quad  \quad+\gamma^s_k\sum_i(\boldsymbol{{\rm{X}}}_k^s)_{i,m}\sum_j(\boldsymbol{{\rm{H}}}^s\boldsymbol{{\rm{X}}}_k^s)_{j,n} (\boldsymbol{{\rm{X}}}_{k,o}^s)_{j,n}(\boldsymbol{{\rm{Q}}}_{k,o}^s)_{n,n} \\
  &\quad \quad \quad \quad+\theta^s_k\sum_i(\boldsymbol{{\rm{X}}}_k^s)_{i,m}\sum_j(\boldsymbol{{\rm{X}}}_k^s)_{j,n}(\boldsymbol{{\rm{T}}}^s\boldsymbol{{\rm{C}}})_{j,n} \\
  & \quad f(\boldsymbol{{\rm{D}}}_k^s)_1 = \alpha^s_k({\boldsymbol{{\rm{X}}}_k^s}^{\rm{T}}\boldsymbol{{\rm{X}}}_k^s\boldsymbol{{\rm{D}}}_k^s{\boldsymbol{{\rm{X}}}_k^s}^{\rm{T}}\boldsymbol{{\rm{X}}}_k^s)_{m,n}\\
  &\quad \quad \quad \quad+\gamma^s_k \sum _i(\boldsymbol{{\rm{X}}}_k^s)_{i,m}\sum_j ((\boldsymbol{{\rm{H}}}^s\boldsymbol{{\rm{X}}}_k^s)_{j,n})^2(\boldsymbol{{\rm{Q}}}_k^s)_{n,n}\\
  &\quad \quad  \quad \quad+\theta_k^s \sum_i(\boldsymbol{{\rm{X}}}_k^s)_{i,m} \sum_j(\boldsymbol{{\rm{X}}}_k^s)^2_{j,n}(\boldsymbol{{\rm{Q}}}_k^s)_{n,n}
  \end{aligned}
\end{equation}

4) For $\boldsymbol{{\rm{D}}}_{k,o}^s$, we also need to pay attention to the $\boldsymbol{{\rm{D}}}_{k,o}^s$ variable of $\boldsymbol{{\rm{Q}}}_{k,o}^s$ in Eq.\ref{eq_globalcommunitydecomposition}. By utilizing the Lagrangian function of $\boldsymbol{{\rm{D}}}_{k,o}^s$, we can derive the iterative update rule:
\begin{equation}\label{eq_dosolver}
  \begin{aligned}
    & \quad \quad \quad \quad (\boldsymbol{{\rm{D}}}_{k,o}^s)_{m,n}=(\boldsymbol{{\rm{D}}}_{k,o}^s)_{m,n}\sqrt{\frac{f(\boldsymbol{{\rm{D}}}_{k,o}^s)_0}{f(\boldsymbol{{\rm{D}}}_{k,o}^s)_1} } \\
    & f(\boldsymbol{{\rm{D}}}_{k,o}^s)_0 = \beta^s_k({\boldsymbol{{\rm{X}}}_{k,o}^s}^{\rm{T}}\boldsymbol{{\rm{M}}}_{k,o}^s\boldsymbol{{\rm{X}}}_{k,o}^s)_{m,n}\\
    &\quad \quad \quad \quad +\gamma^s_k\sum_i(\boldsymbol{{\rm{X}}}_{k,o}^s)_{i,m}\sum_j(\boldsymbol{{\rm{H}}}^s\boldsymbol{{\rm{X}}}_k^s)_{j,n} (\boldsymbol{{\rm{Q}}}_k^s)_{n,n}(\boldsymbol{{\rm{X}}}_{k,o}^s)_{j,n} \\
    & f(\boldsymbol{{\rm{D}}}_{k,o}^s)_1 = \beta^s_k({\boldsymbol{{\rm{X}}}_{k,o}^s}^{\rm{T}}\boldsymbol{{\rm{X}}}_{k,o}^s\boldsymbol{{\rm{D}}}_{k,o}^s{\boldsymbol{{\rm{X}}}_{k,o}^s}^{\rm{T}}\boldsymbol{{\rm{X}}}_{k,o}^s)_{m,n}\\
    &\quad \quad \quad \quad +\gamma^s_k\sum_i(\boldsymbol{{\rm{X}}}_{k,o}^s)_{i,m}\sum_j(\boldsymbol{{\rm{X}}}_{k,o}^s)^2_{j,n}(\boldsymbol{{\rm{Q}}}_{k,o}^s)_{n,n} 
    \end{aligned}
\end{equation} 
 
5) For $\boldsymbol{{\rm{C}}}$, the update ruler can be  through the Lagrangian function of $\boldsymbol{{\rm{C}}}$, we can obtain:
\begin{equation}\label{eq_csolver}
  \begin{aligned}
    \sum_{\boldsymbol{{\rm{S}}}} &\sum_{\boldsymbol{{\rm{K}}}} \theta_k^s(-2{\boldsymbol{{\rm{T}}}^s}^{\rm{T}}\boldsymbol{{\rm{X}}}_k^s\boldsymbol{{\rm{Q}}}_k^s+2{\boldsymbol{{\rm{T}}}^s}^{\rm{T}}\boldsymbol{{\rm{T}}}^s\boldsymbol{{\rm{C}}})=0 \\
    &{\boldsymbol{{\rm{C}}}}_{m,n}=\frac{(\sum_{\boldsymbol{{\rm{S}}}} \sum_{\boldsymbol{{\rm{K}}}} \theta_k^s{\boldsymbol{{\rm{T}}}^s}^{\rm{T}} \boldsymbol{{\rm{X}}}_k^s\boldsymbol{{\rm{Q}}}_k^s )_{m,n}}{(\sum_{\boldsymbol{{\rm{S}}}} \sum_{\boldsymbol{{\rm{K}}}} \theta_k^s {\boldsymbol{{\rm{T}}}^s}^{\rm{T}}{\boldsymbol{{\rm{T}}}^s})_{m,n}} 
  \end{aligned}
\end{equation} 

\subsection{Global Community Detection Algorithm}
Based on the parameter update rules, we give the algorithm \ref{alg_Algorithm of HMCD} to update the parameters of the community detection model, 
and then obtain the global community structure.
\begin{algorithm}[htb]
  \caption{HMCD Algorithm}
  \label{alg_Algorithm of HMCD}
  \renewcommand{\algorithmicrequire}{\textbf{input:}}
  \renewcommand{\algorithmicensure}{\textbf{output:}}
  \begin{algorithmic}[1]
  \REQUIRE User adjacency matrices $\boldsymbol{{\rm{M}}}_k^s$, $\boldsymbol{{\rm{M}}}_{k,o}^s$, the number of communities $K$, term weights $\alpha_k^s$, $\beta_k^s$, $\gamma_k^s$, $\theta_k^s$
  \ENSURE User community structure matrices $\boldsymbol{{\rm{X}}}_k^s$, $\boldsymbol{{\rm{X}}}_{k,o}^s$, $\boldsymbol{{ \rm{C}}}$
  \STATE Normalize each row in adjacency matrix $\boldsymbol{{\rm{M}}}_k^s$, $\boldsymbol{{\rm{M}}}_{k,o}^s$, so that $ \Vert (\boldsymbol{{\rm{M}}}_k^s)i,. \Vert_1 =1$ and $\Vert (\boldsymbol{{\rm{M}}}_{k,o}^s)i,. \Vert _1=1$
  \STATE Normalize each adjacency matrix $\boldsymbol{{\rm{M}}}_k^s$, $\boldsymbol{{\rm{M}}}_{k,o}^s$, so that $ \Vert \boldsymbol{{\rm{M}}}_k^s \Vert_1 =1$ and $\Vert \boldsymbol{{\rm{M}}}_{k,o}^s \Vert _1=1$
  \STATE Initialize user community structure matrices $\boldsymbol{{\rm{X}}}_k^s$, $\boldsymbol{{\rm{X}}}_{k,o}^s$, community-community relation matrices $\boldsymbol{{\rm{D}}}_k^s$, $\boldsymbol{{\rm{D}}}_{k,o}^s$
  \REPEAT
  \FOR{each social network $s$}
    \FOR{each attribute $k$}
      \REPEAT
        \STATE Normalize $\boldsymbol{{\rm{X}}}_k^s$, $\boldsymbol{{\rm{D}}}_k^s$ by $\boldsymbol{{\rm{X}}}_k^s\leftarrow \boldsymbol{{\rm{X}}}_k^s \boldsymbol{{\rm{Q}}}_k^ s$, $\boldsymbol{{\rm{D}}}_k^s\leftarrow (\boldsymbol{{\rm{Q}}}_k^s)^{-1} \boldsymbol{{\rm{D} }}_k^s (\boldsymbol{{\rm{Q}}}_k^s)^{-1}$
        \STATE Update $\boldsymbol{{\rm{X}}}_k^s$ by Eq.\ref{eq_xsolver}
        \STATE Update $\boldsymbol{{\rm{D}}}_k^s$ by Eq.\ref{eq_dsolver}
        \STATE Normalize $\boldsymbol{{\rm{X}}}_{k,o}^s$, $\boldsymbol{{\rm{D}}}_{k,o}^s$ by $\boldsymbol{{\rm{X}}}_{k,o}^s\leftarrow \boldsymbol{{\rm{X}}}_{k,o}^s \boldsymbol{{\rm{Q }}}_{k,o}^s$, $\boldsymbol{{\rm{D}}}_{k,o}^s\leftarrow (\boldsymbol{{\rm{Q}}}_{k,o}^s )^{-1} \boldsymbol{{\rm{D}}}_{k,o}^s (\boldsymbol{{\rm{Q}}}_{k,o}^s)^{-1 }$
        \STATE Update $\boldsymbol{{\rm{X}}}_{k,o}^s$ by Eq.\ref{eq_xosolver}
        \STATE Update $\boldsymbol{{\rm{D}}}_{k,o}^s$ by Eq.\ref{eq_dosolver}
      \UNTIL Eq.\ref{eq_overlappingusercommunitydecomposition} converged
    \STATE Update $\boldsymbol{{\rm{C}}}$ by Eq.\ref{eq_csolver}
    \ENDFOR
  \ENDFOR
  \UNTIL Eq.\ref{eq_globalcommunitydecomposition} converge
  \RETURN community structure matrices $\boldsymbol{{\rm{X}}}_k^s$, $\boldsymbol{{\rm{X}}}_{k,o}^s$, $\boldsymbol{{ \rm{C}}}$
  \end{algorithmic}
\end{algorithm}

\subsection{Algorithm Time Complexity and Convergence Analysis}
\subsubsection{\textbf{Time Complexity}}
This paper designs an algorithm for detecting communities in multiple social networks wherein the parameters are iteratively updated through Algorithm\ref{alg_Algorithm of HMCD}. 
The algorithm's lines (4 to 12) are repeatedly executed over numerous rounds, with the first layer of \textbf{for} statement run $|\boldsymbol{\rm{S}}|$ times, 
and the second layer of \textbf{for} statement run $|\boldsymbol{k}|$ times. 
The primary time consumption is found in the matrix operations within the matrix decomposition. 
To facilitate description, we denote \textbf{the number of users $\boldsymbol{{\rm{U}}}$, $\boldsymbol{{\rm{U}}}^s$ and $\boldsymbol{{\rm{U}}}_o^s$ as $n$, $n_s$ and $n_{s,o}$} respectively. 
Generally, the number of users is significantly greater than the number of communities, $n$,$n_s$,$n_{s,o}$ $\gg$ $ K $.
To be more specific, we take an example of parameters $\boldsymbol{{\rm{X}}}_k^s$ and $\boldsymbol{{\rm{D}}}_k^s$.

The time complexity for the numerator part of $\boldsymbol{{\rm{X}}}_k^s$ update equation is:
${\rm{O}}(2n_s^2K)+{\rm{O}}(2n_sK^2)+{\rm{O}}(n_sKn_{s,o})+{\rm{O}}(n_sK^2)+{\rm{O}}(n_sKn) \approx {\rm{O}}(n_s^2K+nn_sK+n_sn_{s,o}K)$.
The time complexity for the denominator part of $\boldsymbol{{\rm{X}}}_k^s$ update equation is:
${\rm{O}}(2n_sK^2)+{\rm{O}}(2n_s^2K)+{\rm{O}}(2n_s^2K)+{\rm{O}}(2n_sK^2)+{\rm{O}}(n_{s,o}Kn_s)+{\rm{O}}(n_sKn_{s,o}) +{\rm{O}}(1) \approx {\rm{O}}(n_s^2K+n_sn_{s,o}K)$.
Hence, the time complexity for updating the value of $\boldsymbol{{\rm{X}}}_k^s$ is: ${\rm{O}}(n_s^2K+nn_sK+n_sn_{s,o}K)+{\rm{O}}(n_s^2K+n_sn_{s,o}K) \approx {\rm{O}}(n_s^2K+nn_sK+n_sn_{s,o}K)$.
 
The time complexity for the numerator part of $\boldsymbol{{\rm{D}}}_k^s$ update equation is:
${\rm{O}}(n_s^2K)+{\rm{O}}(n_sK^2)+{\rm{O}}(n_{s,o}Kn_s+2n_s)+{\rm{O}}(n_sKn+2n_s)  \approx {\rm{O}}(n_s^2K+n_{s,o}n_sK+nn_sK)$. 
The time complexity for the denominator part of $\boldsymbol{{\rm{D}}}_k^s$ update equation is:
${\rm{O}}(n_sK^2)+{\rm{O}}(K^3)+{\rm{O}}(n_sK^2)+{\rm{O}}(n_sK^2) +{\rm{O}}(n_{s,o}Kn_s+2n_s) +{\rm{O}}(2n_s)\approx {\rm{O}}(n_{s,o}n_sK)$. 
Consequently, the time complexity for updating the value of $\boldsymbol{{\rm{D}}}_k^s$ is:
${\rm{O}}(n_s^2K+n_{s,o}n_sK+nn_sK)+{\rm{O}}(n_{s,o}n_sK) \approx {\rm{O}}(n_s^2K+n_{s,o}n_sK+nn_sK)$.

Similarly, we can obtain the time complexity for updating the value of $\boldsymbol{{\rm{X}}}_{k,o}^s$ is ${\rm{O}}(n_{s,o}^2K)$, 
the time complexity for updating the value of $\boldsymbol{{\rm{D}}}_{k,o}^s$ is ${\rm{O}}(n_{s,o}Kn_s)$, 
and the time complexity for updating the value of $\boldsymbol{{\rm{C}}}$ is ${\rm{O}}(nn_sK)$.
Therefore, the overall time complexity for this portion is:
${\rm{O}}(n_s^2K+nn_sK+n_sn_{s,o}K) + {\rm{O}}(n_s^2K+n_{s,o}n_sK+nn_sK) +{\rm{O}}(n_{s,o}^2K)+{\rm{O}}(n_{s,o}Kn_s)+{\rm{O}}(nn_sK) 
\approx {\rm{O}}(n_s^2K+n_{s,o}^2K+nn_sK+n_sn_{s,o}K)\approx {\rm{O}}(nn_sK+n_sn_{s,o}K)$.
Thus, the time complexity of the algorithm is ${\rm{O}}(ct|\boldsymbol{{\rm{S}}}||\boldsymbol{k}|(nn_sK+n_sn_{s,o}K))$, 
where $ct$ represents the number of iterations in the algorithm.

\subsubsection{\textbf{Algorithm Convergence}}
The method of constructing auxiliary functions\cite{seung2001algorithms} is employed to demonstrate the convergence of the objective function.

This section provides a concise overview of the construction principle of the auxiliary function. 
An auxiliary function must be constructed to satisfy the condition of the given objective function:
\begin{equation}\label{eq_gassistant}
  \left\{
  \begin{aligned}
    &G(x,x') \geq F(x) \quad \quad x \neq x'\\
    &G(x,x')=F(x)  \quad \quad x = x'
  \end{aligned}
  \right.
\end{equation} 

Then, let $x^{t+1}=\arg\min_x G(x,x^t)$. $F(x^{t+1})\leq  G(x^{t+1},\\ x^t) \leq G(x^t,x^t)=F(x^t)$ can be obtained, so this can ensure that the function is monotonicity after each round of value update, and then converges to the extreme point. 
In this paper, the model parameters $\boldsymbol{{\rm{X}}}_k^s$ and $\boldsymbol{{\rm{D}}}_k^s$ are taken as examples, and the convergence proof of the algorithm is given. 
For details, please refer to appendix.

\section{Experiment}
This section conducts experiments to validate the effectiveness of the community detection method. 
Firstly, we present the dataset and experimental setup. 
Subsequently, we describe the evaluation metrics and compare the proposed model, HMCD, with existing works. 
Finally, we present the analysis and discussion of the results.

\subsection{Dataset}
\subsubsection{Synthetic Datasets}
To evaluate the performance of the proposed model, this paper first employs a benchmark framework\cite{bazzi2016generative,multilayerGMpy} to generate synthetic datasets of multilayer networks.
Specifically, a multilayer partition is generated. This partition includes the number of nodes in each layer and the number of layers. It also includes an interlayer dependency tensor. This tensor specifies the desired dependency structure between layers.
Subsequently, for the given multilayer partition, edges in each layer are generated following a degree-corrected block model parameterized by the distribution of expected degrees, a community mixing parameter $\mu \in (0, 1)$. 
The mixing parameter $\mu$ controls the community structure of the network. 
When $\mu=0$, all edges are sampled based on communities. For $\mu=1$ implies that edges are sampled independently. The generated network has no community structure.
The parameter $p \in (0, 1)$ controls the correlation across layers. When $p = 0$ , the community partitions are independent across layers while $p = 1$ indicates an identical community partitions across layers.

\subsubsection{Real-World Datasets}
This paper also uses a dataset of multiple social networks, including Twitter, Instagram, and Tumblr. Twitter is a short-text-based social network. 
Instagram is a photo-sharing-based social network. 
Tumblr is a multimedia blog-based social network that supports a wide range of rich media, such as pictures and videos.

\begin{figure}[htb]
  \centerline{\includegraphics[width=0.35\textwidth]{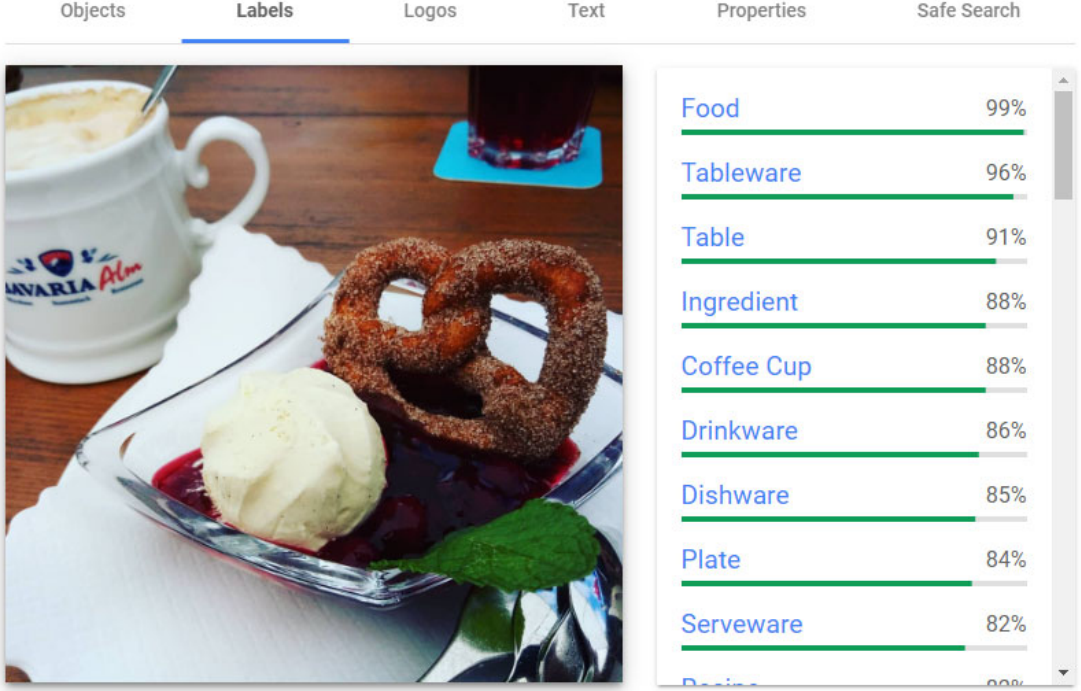}}
  \caption{Example of the photo posted}
  \label{fig_image2text}
\end{figure}

The dataset contains user account information that was collected from About.me, a website that allows users to link multiple online identities through their business cards.
The total number of users is 4716. The user accounts include 2195 Twitter users, 2170 Instagram users, and 351 Tumblr users.
Based on these users, user content data is crawled from the connected social networks respectively. 
The users' content data spans from January 2017 to January 2018, where the non-text type data, such as images, are converted into word tags by Google's deep learning-based image annotation model\footnote[2]{https://cloud.google.com/vision}. 
Fig.\ref{fig_image2text} shows a photo posted on Instagram, for which we used the annotation model to generate tags.
The attribute information of users' topology is created by analyzing their followees and followers on social media networks.
After preprocessing, we filter out some users to keep the connectivity in each social network.

The statistics of the dataset are shown in Tab.\ref{tab_number_user} and Tab.\ref{tab_number_content}. 
Each element in Tab.\ref{tab_numble_user} is the number of identical users corresponding to two social networks. 
1223 users have both Twitter and Instagram accounts, 10 users have both Twitter and Tumblr accounts, 25 users have Instagram and Tumblr accounts.
In addition, 6 identical users link to all three networks. The number of users' content and relation data is shown in Tab.\ref{tab_number_content}.
\begin{table}[ht]
\caption{The Number of Users Across Social Networks}
\begin{center}
\resizebox{0.3\textwidth}{!}{
\begin{tabular}{c|c|c|c}
\hline
&\textbf{Twitter}& \textbf{Instagram}& \textbf{Tumblr} \\
\hline
\textbf{Twitter}&2,195& 1,223 &10\\
\hline
\textbf{Instagram}&-& 2,170 &25\\
\hline
\textbf{Tumblr}&-&- & 351 \\
\hline
\end{tabular}}
\label{tab_number_user}
\end{center}
\end{table}
\begin{table}[ht]
\caption{The Number of Contents}
\begin{center}
  \resizebox{0.3\textwidth}{!}{
\begin{tabular}{c|c|c|c}
\hline
&\textbf{Twitter}& \textbf{Instagram}& \textbf{Tumblr} \\
\hline
\textbf{\#content}&721,612 & 345,221 &143,088 \\
\hline
\textbf{\#relation}&67,270 & 36,363 &1,150 \\
\hline
\end{tabular}}
\label{tab_number_content}
\end{center}
\end{table}

\subsection{Setup}
\begin{figure*}[htb]
  \centering
  \begin{minipage}[b]{0.82\textwidth}
    \centering
    \subfloat[mod(G)]{
      \includegraphics[width=0.33\textwidth]{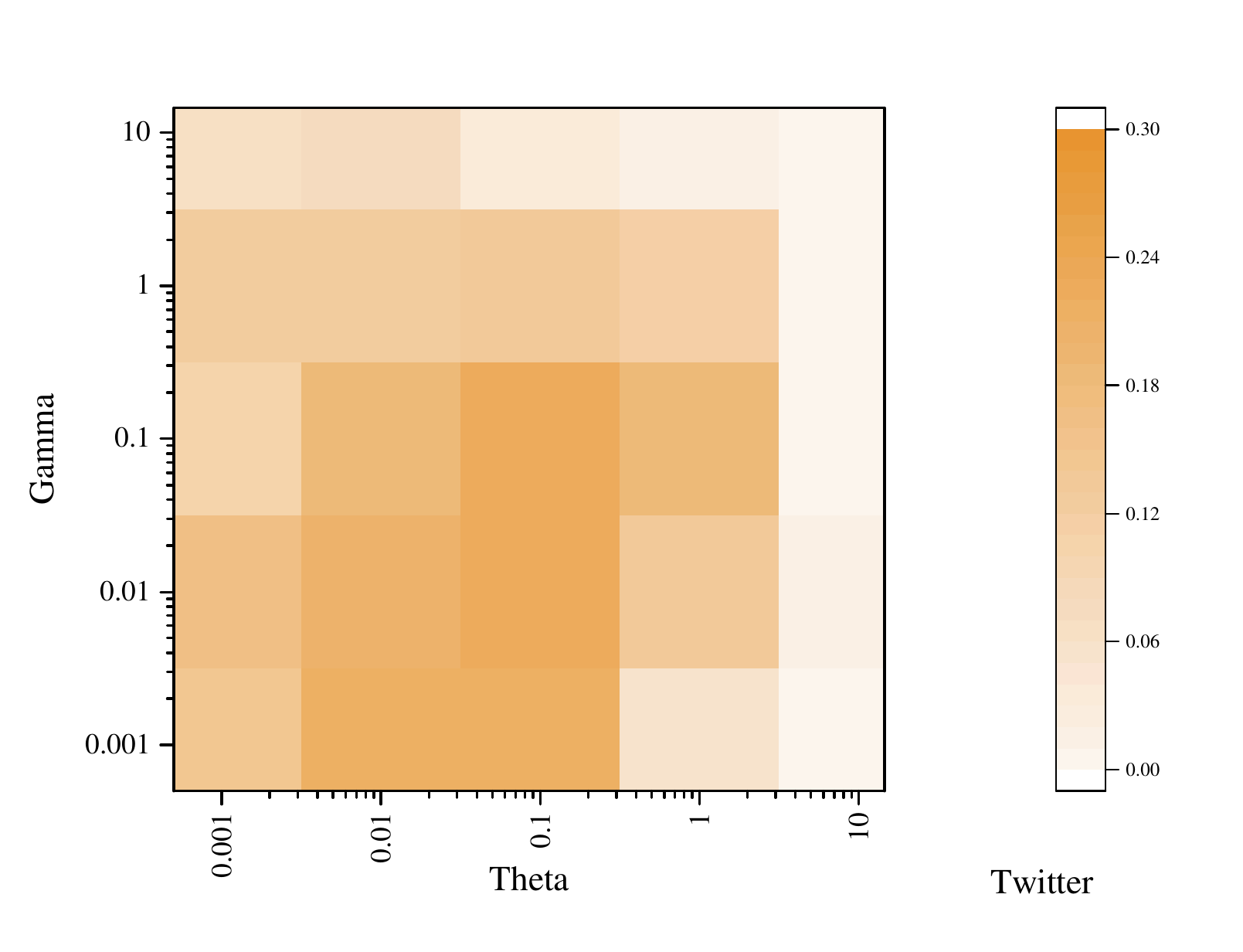} 
      \hspace*{0.02cm}
      \includegraphics[width=0.33\textwidth]{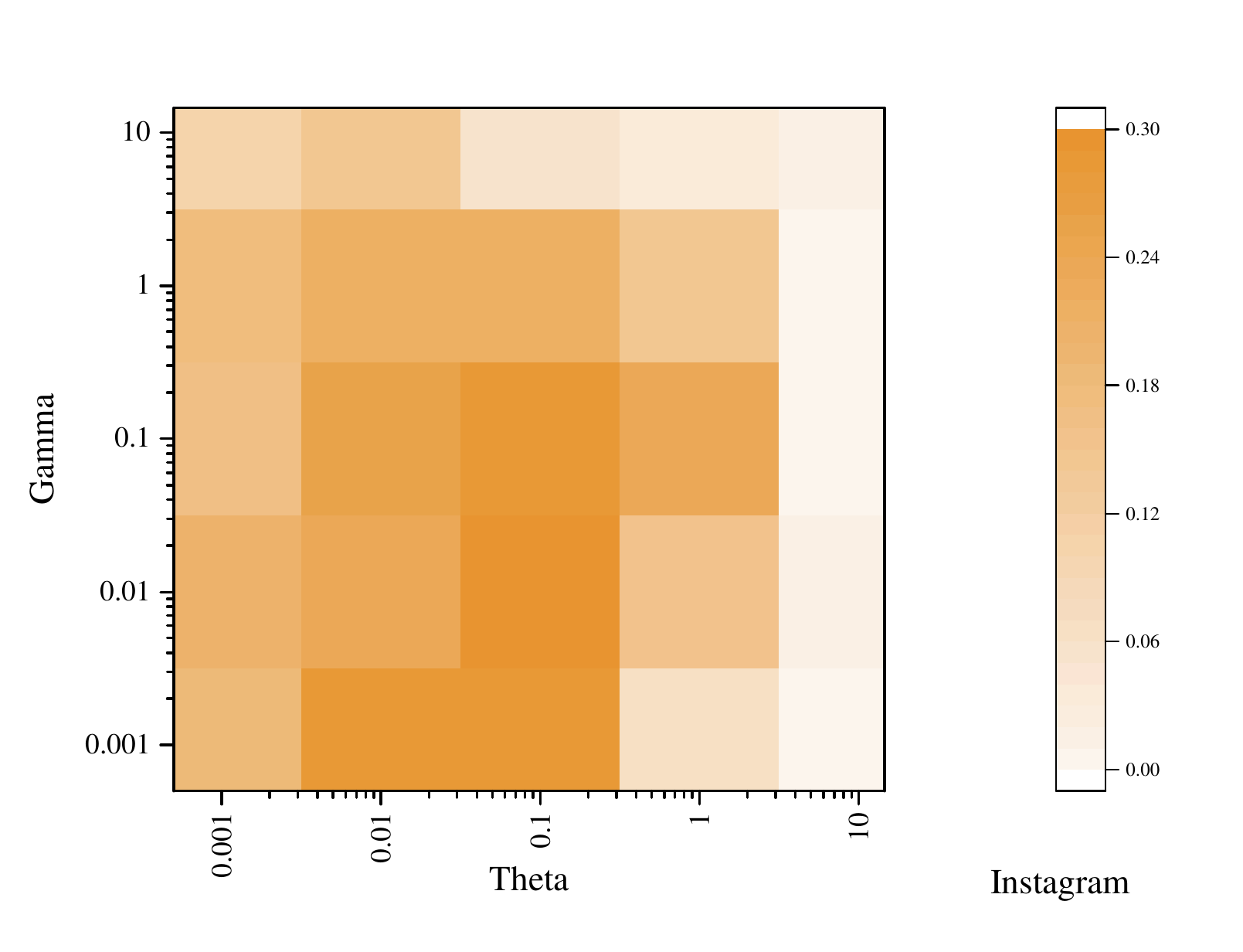}
      \hspace*{0.02cm}
      \includegraphics[width=0.33\textwidth]{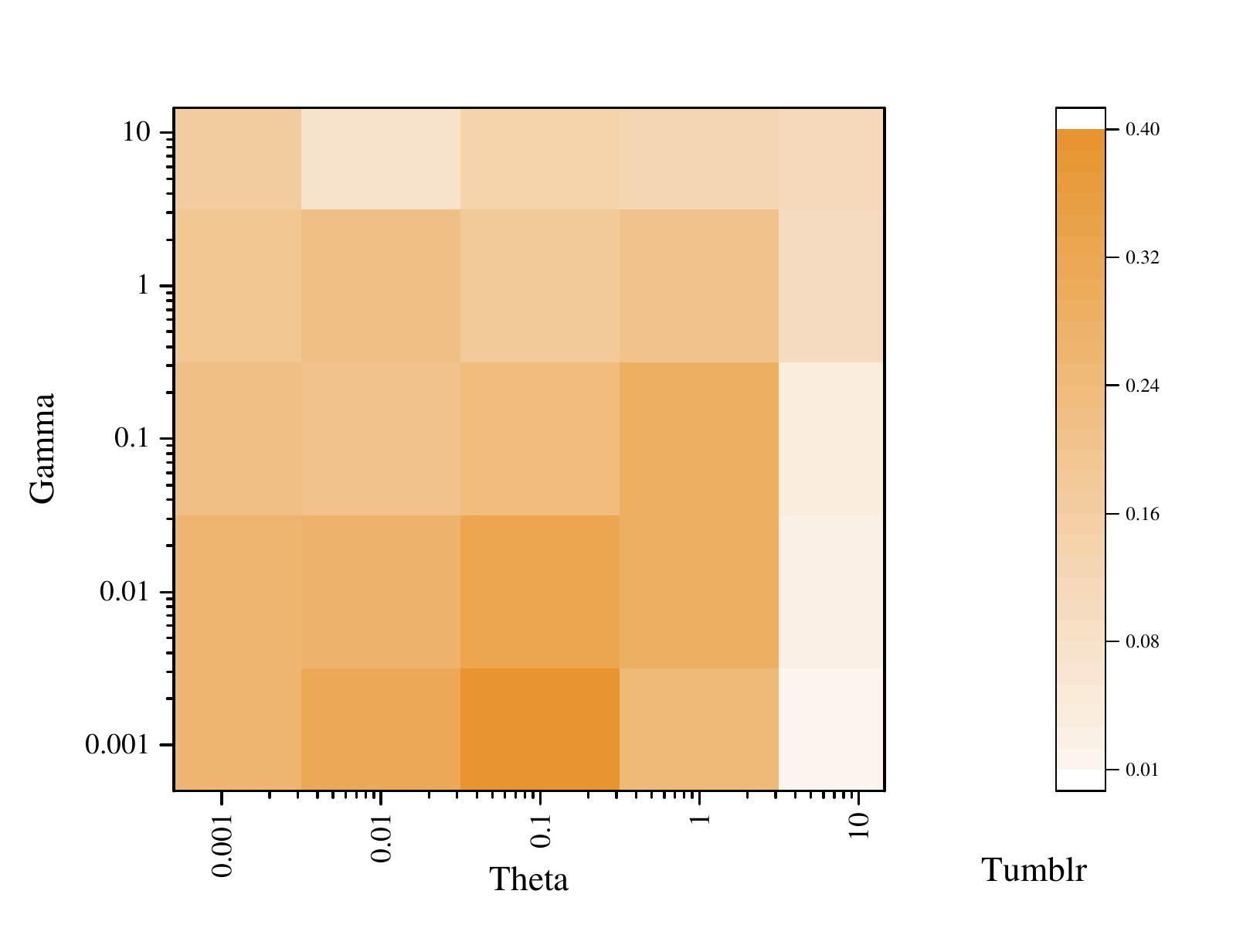} 
      \label{fig_globalcommunitymodularityexp}
  }
  \end{minipage}
  \begin{minipage}[b]{0.82\textwidth}
    \centering
    \subfloat[mod\textsuperscript{o}(G)]{
      \includegraphics[width=0.33\textwidth]{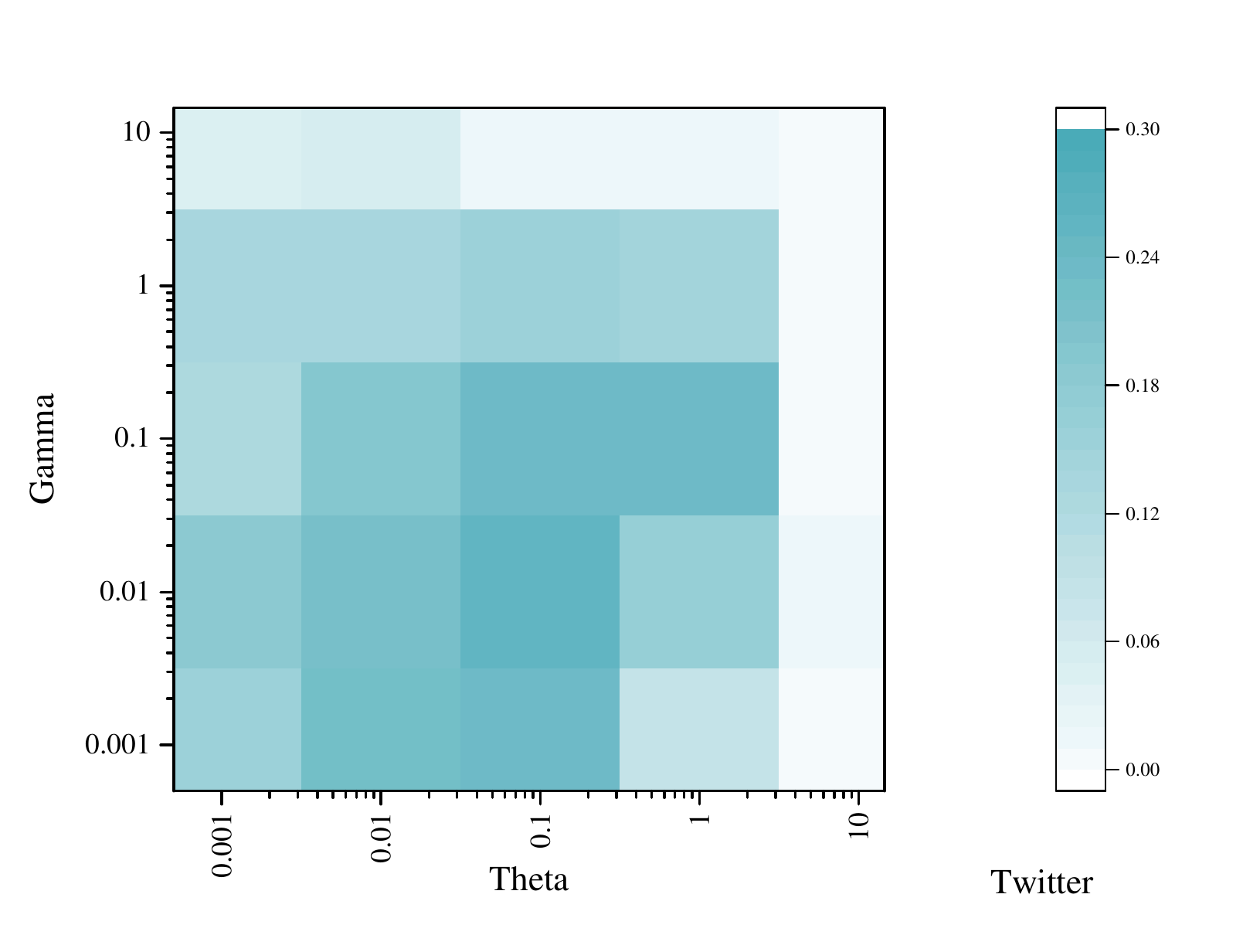}
      \hspace*{0.02cm}
      \includegraphics[width=0.33\textwidth]{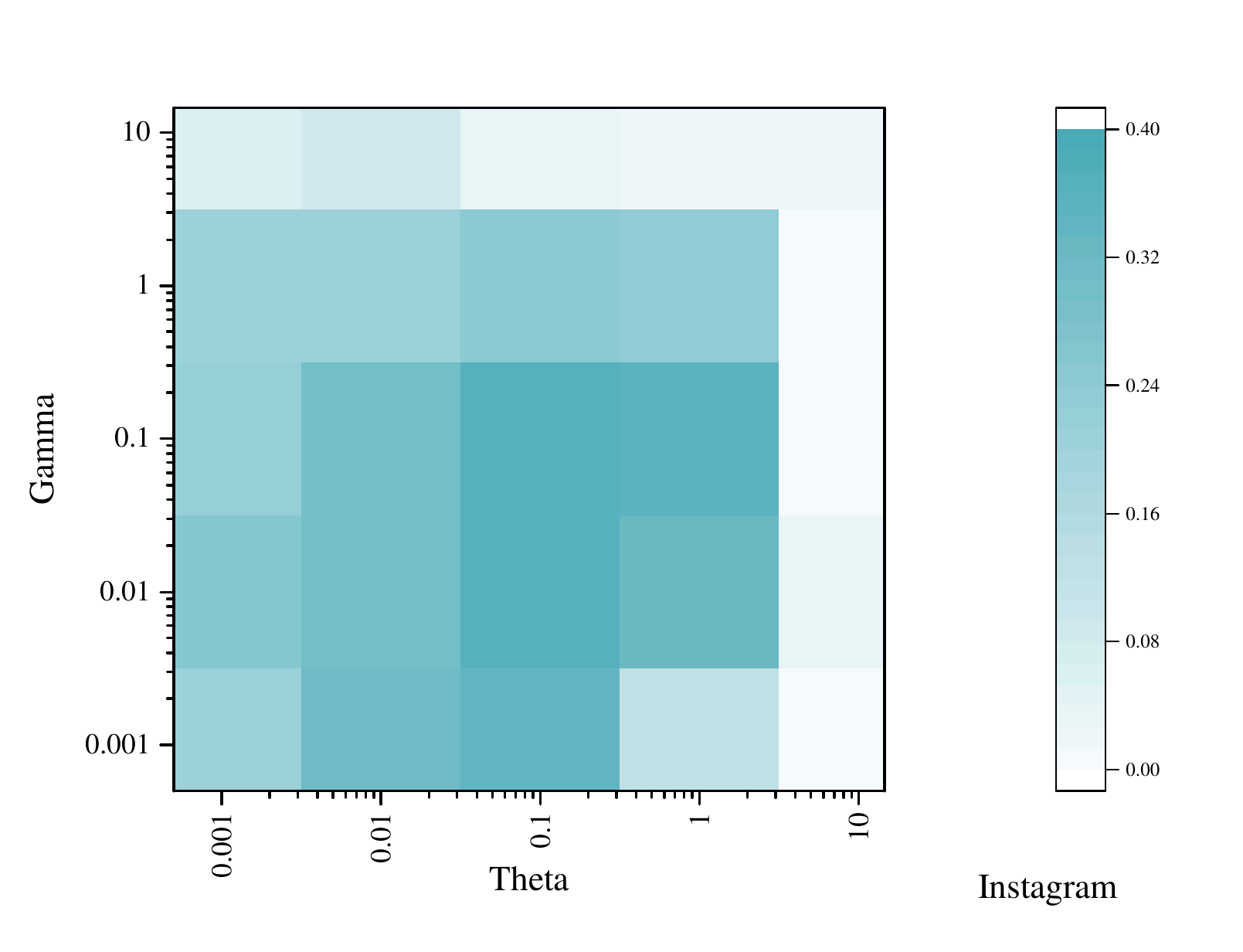}
      \hspace*{0.02cm}
      \includegraphics[width=0.33\textwidth]{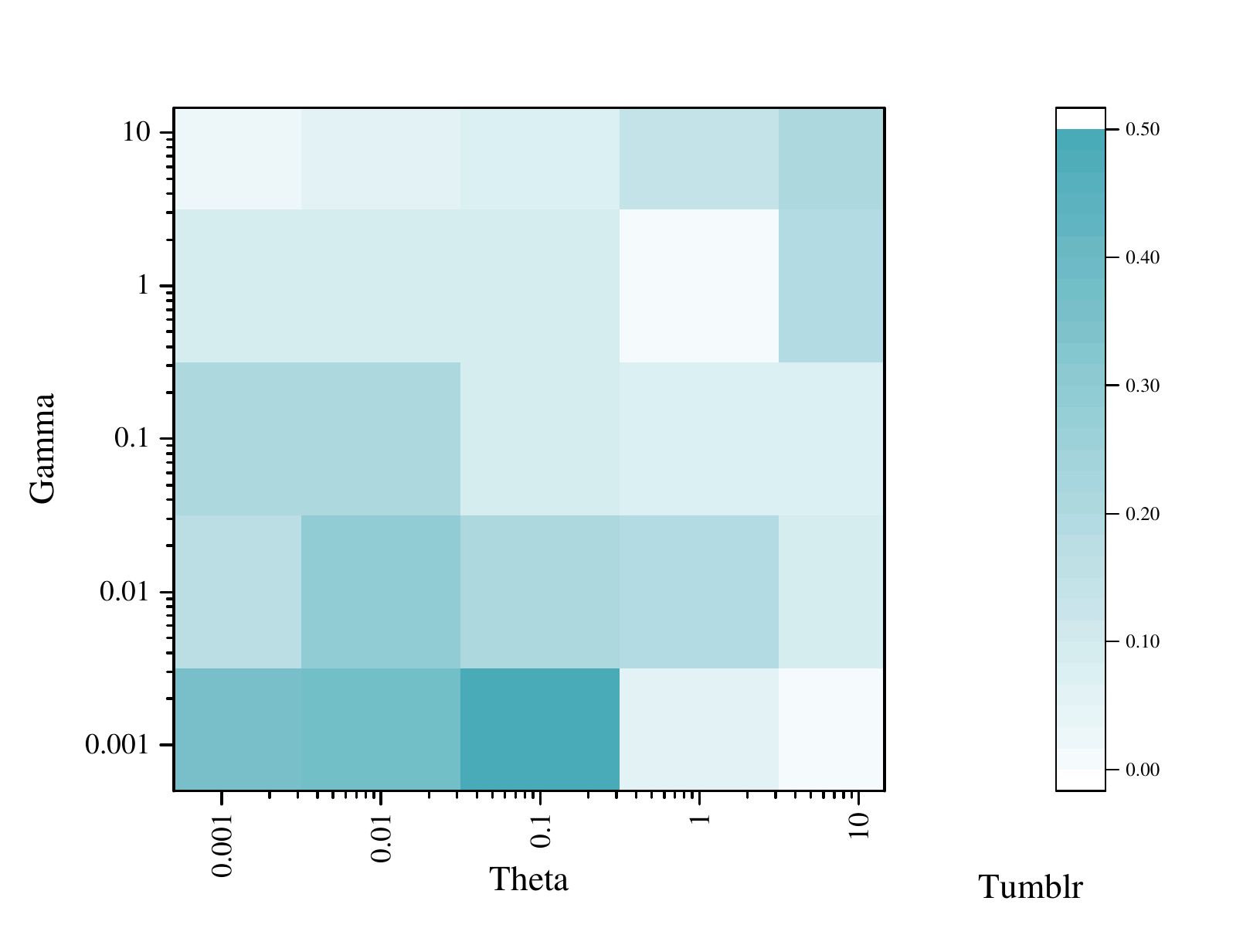} 
      \label{fig_overlappingusersglobalcommunitymodularityexp}
  }
  \end{minipage}
  \caption{
  Experiments on combinations of the parameters $\gamma_k^s$, $\theta_k^s$. We take values of $\gamma_k^s$ and $\theta_k^s$ from $[0.001,0.01,0.1,1,10]$. 
  The focal points of experiments are metrics mod(G), mod\textsuperscript{o}(G), which are modularity between users and modularity between overlapping users based on the different user types within the community.
  The metrics are detailedly described in section\ref{section_communityquality}.
  The experiments primarily illustrate the effects on the topology attribute when the number of community divisions is 40 in the real-world dataset. 
  Evidently discernible from the figure, the model proposed in this paper demonstrates better performance during $\theta_k^s$ is about 0.1 and $\gamma_k^s$ ranges from 0.001 to 0.1.
  }
  \label{fig_parametersexp}
\end{figure*}

In this model, the number of communities is determined by experimental results.
We randomly initialize the element values in community structure matrix $\boldsymbol{{\rm{X}}}$ and community-community relation matrix $\boldsymbol{{\rm{D}}}$ within the range of 0 to 1.
The weights $\alpha_k^s$ and $\beta_k^s$ are mainly used to represent the importance of social networks. 
In this paper, each social network has the same significance, and we set all values to $1$, that is $\alpha_k^s=\beta_k^s=1$. 
$\gamma_k^s$, $\theta_k^s$ can affect the ability of HMCD to integrate the community structure of each social network. Moreover, each social network has the same $\gamma_k^s$, $\theta_k^s$.
The larger the value of $\theta_k^s$ is relative to the value of $\alpha_k^s$, $\beta_k^s$, 
the more the global community ignores the intrinsic community feature in each social network, 
and the smaller the value is, the worse the global community integrates the community information in each social network. Similarly, the $\gamma_k^s$ represents the significance weight of the overlapping users.
The value of $\gamma_k^s$ controls the contribution of overlapping community.
To facilitate the comparison with other works, we take values of $\gamma_k^s$ and $\theta_k^s$ from $[10^{-a},...,0.01,0.1,1,...,10^b]$\cite{liu2013multi,he2014comment}, and the parameters with the best experimental results are selected as model parameters. 
As shown in Fig.\ref{fig_parametersexp}, we conduct an experiment to discern the appropriate combinations of parameters $\gamma_k^s$, $\theta_k^s$.
Based on our analysis of the experimental results, we use $\gamma_k^s=0.01$, $\theta_k^s=0.1$ in the experimental part of this paper.

\subsubsection{Synthetic Dataset}
\begin{figure}[htb]
  \centerline{\includegraphics[width=0.38\textwidth]{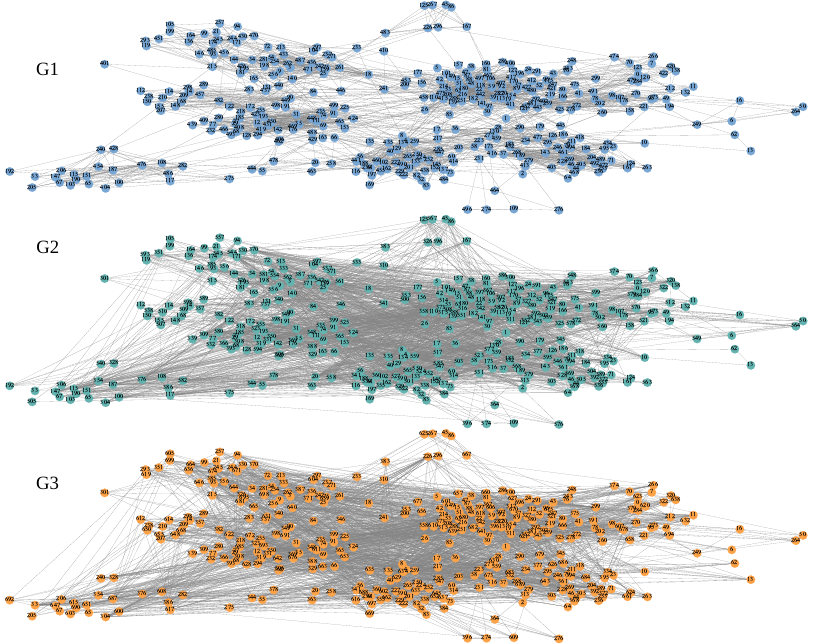}}
  \caption{We create synthetic data of multiple social networks by fixing $\theta=0.3$, $p=0.7$, and using the default values of $\mu=0.1$, $k\_min=5$, $k\_max=70$, $t_k=-2$ in the generative benchmark. The number of communities is set to 20.}
  \label{fig_multiplegrahps}
\end{figure}

We use a multilayer network benchmark to establish three-layer networks, with each layer consisting of 400 nodes. 
In an attempt to emulate the scenario of multiple social networks, we perceive each layer as a separate social network, denoted as G1, G2, and G3. 
Initially, we randomly select 100 nodes and consider them to be overlapping users who have accounts across all networks. 
Subsequently, we select a network, such as G1, and randomly sample 100 nodes from the remaining nodes, removing their alignment with G2. 
These nodes are regarded as users with accounts in both G1 and G3. 
Then, we randomly select 100 nodes from the remaining nodes, eliminating their alignment with G3, and consider these nodes as users with accounts in both G1 and G2. 
Next, we merely need to randomly select 100 nodes from the remaining nodes in G2 and delete their alignment with G1 to identify users who have accounts in both G2 and G3. 
Ultimately, the remaining 100 nodes in each layer are viewed as users who only have accounts in their respective network. 
Thus, we obtain partially aligned multiple social networks, where 100 users have accounts across all G1, G2, and G3 networks, 
100 users only have accounts in either both G1 and G2, both G1 and G3, or both G2 and G3, and 100 users are only in one of the networks, either G1, G2, or G3.
The structure of synthetic multiple networks is shown in Fig.\ref{fig_multiplegrahps}.
This process enables us to create a realistic simulation of partially aligned multiple social networks.

\subsubsection{Real-World Dataset}
For the topology attributes, this paper uses a $0$ or $1$ value to represent whether there is a followee or follower relationship between users in each social network ,and then constructs the adjacency matrix.
In terms of content attributes,  the model can calculate the similarity between users. 
This results in a denser distribution of value elements in the adjacency matrix compared with the adjacency matrix constructed from topology attributes.
Therefore, we first use the topic model\cite{zhu2021understanding} to obtain the user's topic feature vector.
Then, we use JS divergence (Jensen-Shannon divergence) to measure feature similarity between users, and use the deformed JS divergence\cite{rus2013similarity} ($10^{-JS})$ to convert the value to $(0, 1]$, the larger the value, the higher the similarity. 
Finally, we set the value to $0$ if there is no connection between users and the content similarity is less than $0.7$. 
It has been shown that there is a high likelihood of interaction occurring between two users if they have a topological connection or a content similarity greater than 0.7.

\subsection{Evaluation Metrics}
This paper will conduct experiments from two perspectives.
One aspect involves evaluating community quality, while the other focuses on assessing the degree of community fusion.

\subsubsection{Community Quality}
This section presents several metrics of community quality employed in the experiments.

\textbf{Modularity}: 
First, this paper uses the Modularity\cite{newman2004finding} to measure community quality. 
Its form is shown as Eq.\ref{eq_modularity}. 
Modularity can be used not only to evaluate the community quality over the topology attributes but also to evaluate the community quality over the content attributes.
\begin{equation}\label{eq_modularity}
  {\rm{Q}}_{\rm{mod}}=\frac{1}{\rm{W}}\sum_{ij}(w_{ij}-\frac{e_i^{out}e_j^{in}}{\rm{W}})\delta_{ij} 
\end{equation} 
where $\rm{W}$ represents the sum of all weighted edges of the network. 
$w_{ij}$ represents the weight from node $i$ to node $j$. 
$e_i^{out}$ represents the sum of weights of the outgoing edges from node $i$. 
$e_j^{in}$ represents the sum of weights of the incoming edges from node $j$. 
$\delta_{ij}$ indicates whether node $i$ and node $j$ are in the same community.

\textbf{Compactness}: 
The compactness metric\cite{creusefond2015finding} for multiple social networks is defined to measure the intensity of information spread among community members. 
The spread of information within the community is significant across multiple social networks\cite{das2021deployment}.
Specifically, we assume that information passes through a connecting edge. Consequently, the smaller the diameter of the community, the less time it takes for the information to propagate from one end to the other.
The greater the number of edges in the community, the more conducive it is to the diffusion of information throughout the community.
This paper takes the longest length of the shortest connection path between the community members as the community's diameter. 
It then divides the number of edges connecting with overlapping users in the community by the community's diameter to calculate the compactness for the multiple social networks.
In this way, we evaluate the ability of information to diffuse within the community across multiple social networks.
Its specific form is shown in Eq.\ref{eq_compactness}.
\begin{equation}\label{eq_compactness}
  {\rm Q_{comp}} = \sum_c \frac{E(c)}{{diam(c)}}
\end{equation} 
where $E(c)$ denotes the number of edges connected to overlapping users in the community $c$. 
$diam(c)$ represents the longest path length among all the shortest connected paths between nodes in the community.

\textbf{Density}:
The content density\cite{chakraborty2017metrics} metric is designed for multiple social networks to measure the content similarity between members of the community. 
Specifically, we will calculate the average similarity of content between the user and the overlapping user in each community ,and then combine the average similarity of all communities to obtain the density value. 
A larger value represents the more similarities among members of the community with overlapping users over content attributes. 
The specific form of the content density metric is shown in Eq.\ref{eq_density}.
\begin{equation}\label{eq_density}
  {\rm Q_{dens}}= \frac{1}{K} \sum_c \frac{1}{{| N(c)|}} \sum_{i,j \in N(c)} sim(i,j)
\end{equation}
where $N(c)$ represents the user pairs within the community $c$, including overlapping users with overlapping users and non-overlapping users with overlapping users.
$sim(i,j)$ represents the content similarity between node $i$ and node $j$. $K$ is the number of community detection.

\subsubsection{Community Fusion}
This paper uses the normalized mutual information (NMI)\cite{chakraborty2017metrics} to measure the similarity between the global community and the intrinsic community in each social network. 
This metric can illustrate the ability of the model to fuse different social network communities under the current quality of community division, and its specific form is shown in Eq.\ref{eq_NMI}.
\begin{equation}\label{eq_NMI}
  \begin{aligned}
  &\quad \quad \quad  \quad NMI(c,c_k^s) =\frac{MI(c,c_k^s)}{\sqrt{H(c)H(c_k^s)}}\\
  &MI(c,c_k^s) = \sum_i \sum_j P(c_i,c_{k,j}^s)log\frac{P(c_i,c_{k,j}^s)}{P({c_i})P(c_{k,j}^s)}
  \end{aligned}
\end{equation}

where $H(c)=P(c)log(\frac{1}{P(c)})$,$H(c_k^s)=P(c_k^s)log(\frac{1}{P(c_k^s)})$,
$P(c_i,c_{k,j}^s)=\frac{|c_i \cap c_{k,j}^s|}{N^s}$, $P(c_i)=\frac{N^s(c_i)}{N^s}$, $P(c_{k,j}^s)=\frac{|c_{k,j}^s|}{N^s}$, 
$c_i$ represents the user set of the $i$th global community, 
and $c_{k,j}^s$ represents the user set of the $j$th community over the attribute $k$ in the social network $s$. 
$N^s$ represents the number of users in the social network $s$. 
$N^s(c_i)$ represents the number of users of $i$th global community in social network $s$.
\subsection{Baselines}
To demonstrate the effectiveness of the proposed method, we choose different baseline methods as follows. Apart from MCD, all methods are deployed with the alignment matrix described in this paper. This allows the methods to be applied in multiple social networks.

\textbf{MultiNMF}\cite{liu2013multi}: This model utilizes the consistency factor to approximate the clustering results of each view and employs the standard asymmetric non-negative matrix factorization form to fuse the clustering results from different views. We use one of two factorization factors as community structure matrix.

\textbf{uniCDMN}\cite{nguyen2015community}: The model uses a symmetric non-negative matrix tri-factorization form to fuse communities across social networks through a common factor in the model.

\textbf{coNMTF}\cite{yang2018community}: The model uses the symmetric non-negative matrix tri-factorization form to detect community in each social network. These communities can affect each other in the model. After the community detection in each social network is finally obtained, they are merged to generate the global community.

In the experiments, MultiNMF, uniCDMN, and coNMTF can obtain the community structure in every single social network and the global community for all social networks.

\textbf{coNMF}\cite{he2014comment}: This model employs the standard asymmetric non-negative matrix factorization form, which may integrate the clustering results from different user views into pairs of constraints, allowing them to interact and realize complementary in each view. We use one of two factorization factors which is in pairs of constraints as community structure matrix.

\textbf{MCD}\cite{cao2020mutual}: The model uses spectral clustering to detect the communities in each social network. The community structure between different networks constitutes pairs of constraint terms and, based on the overlapping users' consistency across social networks, adjusts the community structure in each social network.

coNMF and MCD can only obtain the community in every single social network. In the experiment, we get the global community for multiple social networks by the method in\cite{yang2018community} and compare it with our model in this paper.

\textbf{HMCD}: A community detection model for heterogeneous multiple social networks designed in this paper.
\subsection{Results Analysis on Synthetic Dataset}
This section evaluates the model performance based on a synthetic dataset. 
Because the synthetic dataset only possesses topology attributes, our analysis is constrained to these specific features. 
Our specific focus is on modularity, which we further categorize as modularity between users and modularity between overlapping users, depending on the different user types within the community. 
The modularity between overlapping users solely measures the community quality of overlapping users. 
These two types of modularity are denoted as \textbf{mod} and \textbf{mod\textsuperscript{o}}, respectively. 
This paper explores a scenario that encompasses multiple social networks. 
The concept of modularity can be classified into two categories: modularity specific to the local community within a single social network (denoted as \textbf{mod(L)}) and modularity specific to the global community across social networks (denoted as \textbf{mod(G)}). 
The experiment measures four types of modularity: \textbf{mod(L)}, \textbf{mod(G)}, \textbf{mod\textsuperscript{o}(L)} and \textbf{mod\textsuperscript{o}(G)}.

In the previous section, we also established evaluation metrics for topology attributes. 
In experiments, we use these metrics specific to global community structure across social networks.
The metric is denoted as ``\textbf{comp}'' for compactness.

\begin{figure}[htb]
  \centering
  \begin{minipage}[b]{0.24\textwidth}
    \centering
    \subfloat[all users]{
      \includegraphics[width=1.0\textwidth]{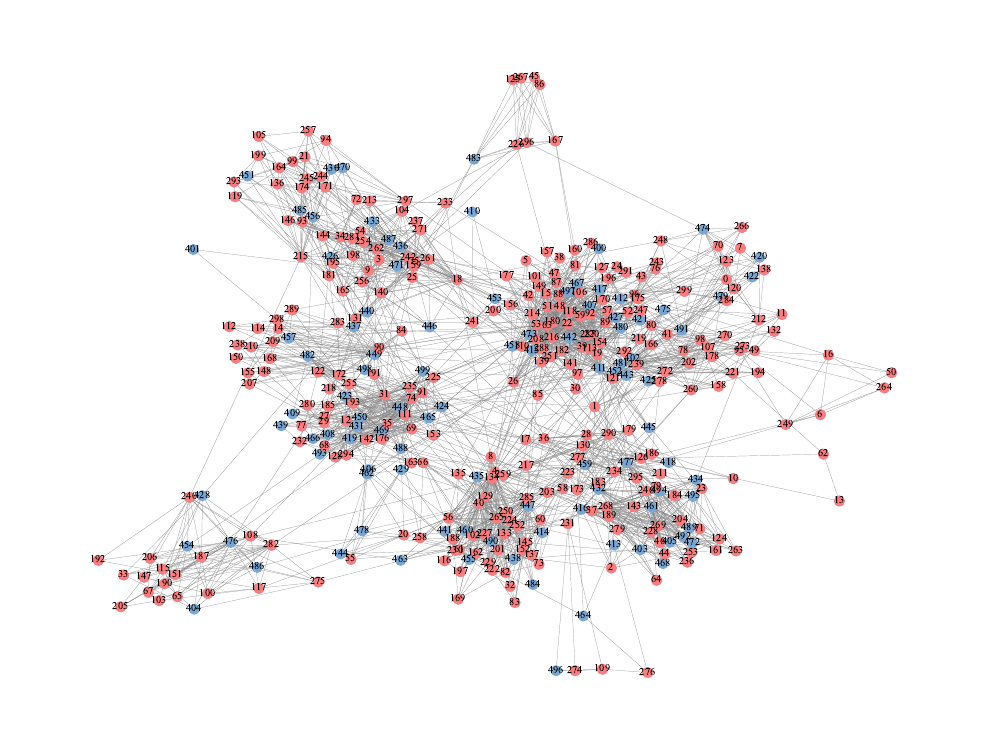} 
      \label{fig_showdetectioncom}
  }
  \end{minipage}
  \begin{minipage}[b]{0.24\textwidth}
    \centering
    \subfloat[only overlapping users]{
      \includegraphics[width=1.0\textwidth]{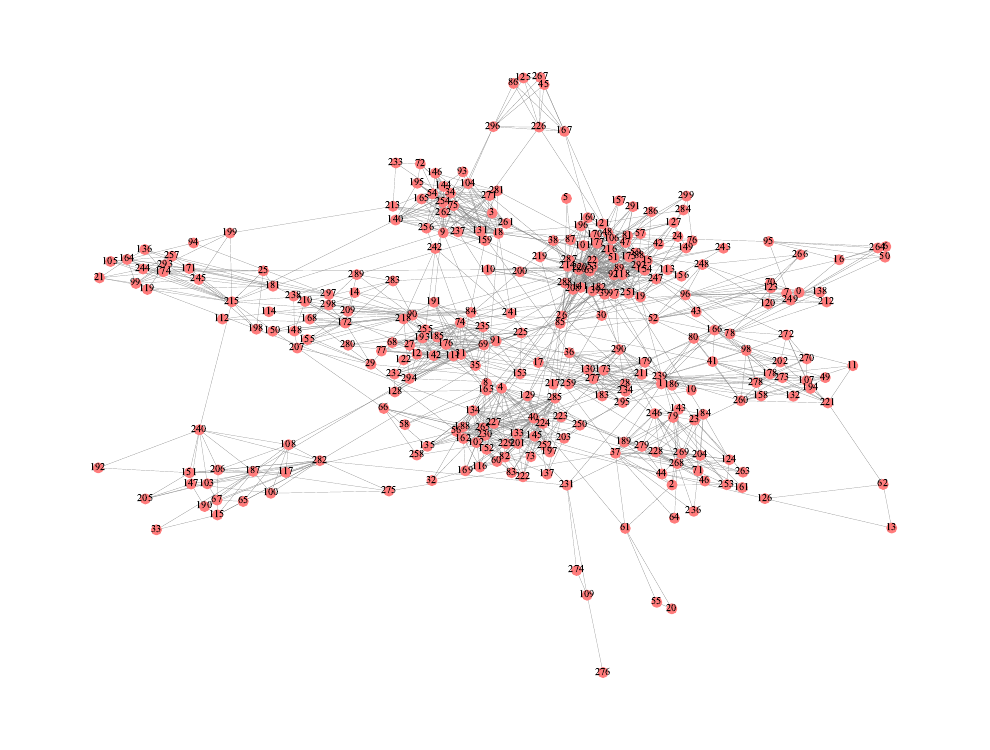}
      \label{fig_showdetectioncom_o}
  }
  \end{minipage}
  \caption{This figure illustrates community detection in our work. Our model, HMCD, is capable of extracting community structures among overlapping users.
  The illustration on the left delineates the network topology of G1, where the red nodes represent the overlapping users within G1. 
  The right side depicts the network topology as perceived from the perspective of the overlapping users in G1.
  }
  \label{fig_overlappingcommunitydetection}
\end{figure}

The experimental result given in Tab.\ref{tab_Synthetic_com_exp} for the community quality shows that our model HMCD performs better than others.
This suggests that our model can efficaciously discover the community integration among overlapping users, while simultaneously upholding a high caliber of community between regular users. 
The network topology ,based on overlapping users in synthetic data, is shown in Fig.\ref{fig_overlappingcommunitydetection}.

\begin{table}[htbp]
  \centering 
  \caption{Comparison of Community Modularity and Compactness Based on Synthetic Dataset} 
  \label{tab_Synthetic_com_exp}
  \resizebox{0.45\textwidth}{!}{%
  \begin{tabular}{ccccccc}
  \toprule
  \multirow{2}{*}{Network}   & \multirow{2}{*}{Method} & \multicolumn{5}{c}{$K=20$}                                                                \\ \cmidrule(l){3-7} 
                             &                         & mod(L)                  & mod\textsuperscript{o}(L)& mod(G)                  & mod\textsuperscript{o}(G)& comp                      \\ \midrule
  \multirow{6}{*}{G1}        & MultiNMF                & 0.547 ±.035             & 0.532 ±.032              & 0.303 ±.052             & 0.237 ±.049              & 562 ±134              \\ 
                             & coNMF                   & 0.564 ±.023             & 0.505 ±.026              & 0.302 ±.019             & 0.223 ±.010              & 519 ±046              \\ 
                             & uniCDMN                 & 0.445 ±.015             & 0.348 ±.017              & 0.445 ±.015             & 0.348 ±.016              & 843 ±034               \\ 
                             & coNMTF                  & 0.448 ±.043             & 0.477 ±.063              & 0.317 ±.049             & 0.303 ±.015              & 573 ±102              \\ 
                             & MCD                     & 0.656 ±.022             & 0.627 ±.017              & 0.343 ±.091             & 0.291 ±.008              & 612 ±185              \\ 
                             & HMCD                    & \textbf{0.678 ±.012}    & \textbf{0.651 ±.017}     & \textbf{0.522 ±.028}    & \textbf{0.456 ±.010}     & \textbf{933 ±068}      \\ \midrule
  \multirow{6}{*}{G2}        & MultiNMF                & 0.576 ±.029             & 0.578 ±.015              & 0.339 ±.027             & 0.305 ±.023              & 627 ±074              \\ 
                             & coNMF                   & 0.621 ±.029             & 0.590 ±.036              & 0.317 ±.013             & 0.267 ±.033              & 552 ±042               \\ 
                             & uniCDMN                 & 0.485 ±.021             & 0.415 ±.024              & 0.485 ±.021             & 0.415 ±.024              & 863 ±079             \\ 
                             & coNMTF                  & 0.551 ±.161             & 0.562 ±.153              & 0.357 ±.102             & 0.333 ±.094              & 611 ±185              \\ 
                             & MCD                     & 0.650 ±.026             & 0.640 ±.032              & 0.435 ±.058             & 0.396 ±.055              & 766 ±129              \\  
                             & HMCD                    & \textbf{0.685 ±.044}    & \textbf{0.667 ±.045}     & \textbf{0.519 ±.031}    & \textbf{0.470 ±.024}     & \textbf{896 ±058}      \\ \midrule
  \multirow{6}{*}{G3}        & MultiNMF                & 0.563 ±.014             & 0.528 ±.011              & 0.392 ±.095             & 0.293 ±.040              & 572 ±135      \\ 
                             & coNMF                   & 0.606 ±.053             & 0.554 ±.065              & 0.333 ±.016             & 0.250 ±.007              & 569 ±040                 \\ 
                             & uniCDMN                 & 0.482 ±.020             & 0.390 ±.017              & 0.482 ±.020             & 0.390 ±.017              & 924 ±069                 \\ 
                             & coNMTF                  & 0.593 ±.041             & 0.590 ±.026              & 0.385 ±.056             & 0.345 ±.074              & 684 ±130                \\ 
                             & MCD                     & 0.643 ±.010             & 0.601 ±.017              & 0.367 ±.017             & 0.316 ±.027              & 620 ±077               \\ 
                             & HMCD                    & \textbf{0.712 ±.002}    & \textbf{0.668 ±.006}     & \textbf{0.580 ±.024}    & \textbf{0.517 ±.023}     & \textbf{979 ±074}   \\ 
  \bottomrule
  \end{tabular}%
  }
\end{table}

\begin{figure*}[htb]
  \centering
  \begin{minipage}[b]{0.83\textwidth}
    \centering
    \subfloat{
      \includegraphics[width=0.30\textwidth]{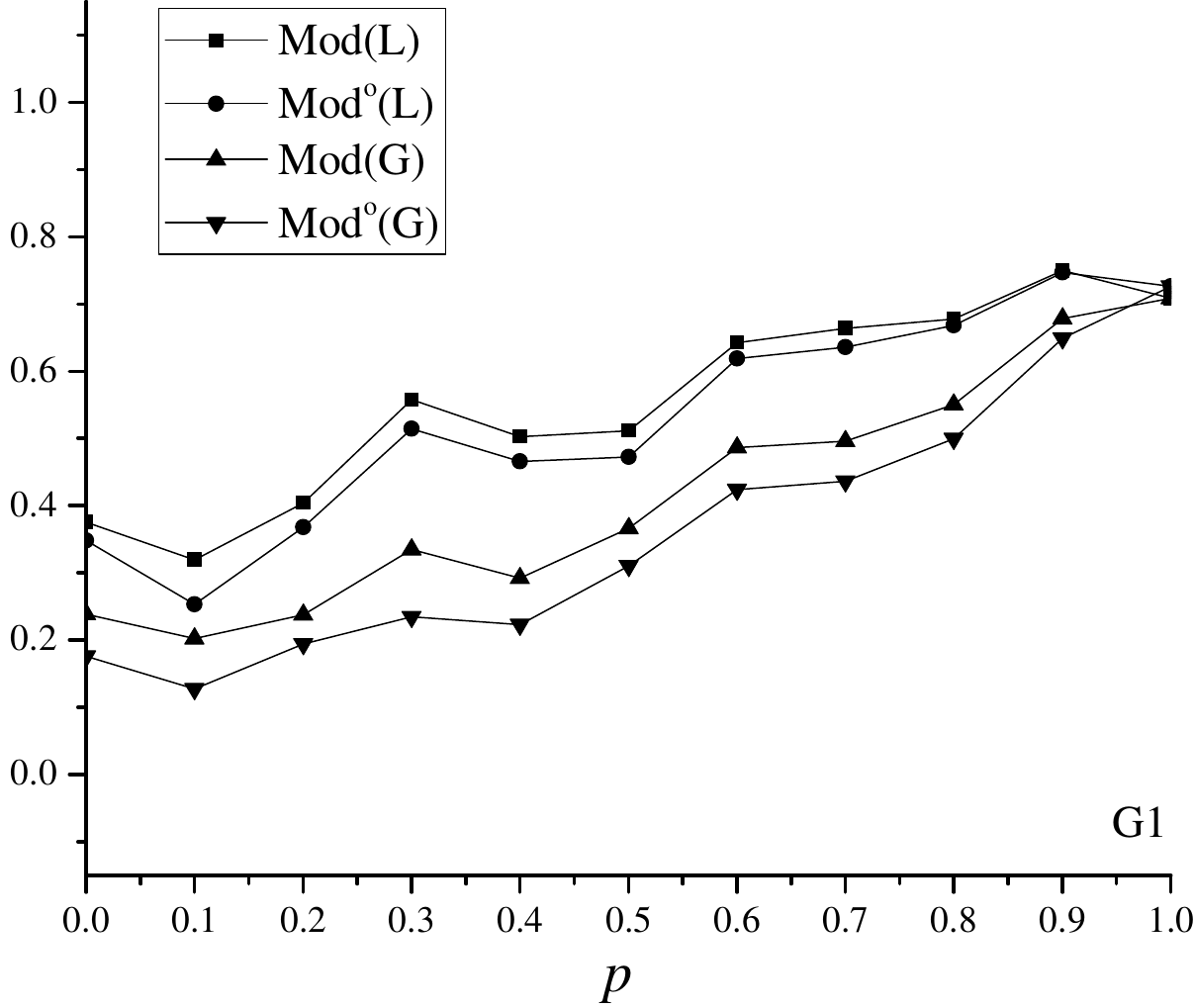} 
      \hspace*{0.2cm}
      \includegraphics[width=0.30\textwidth]{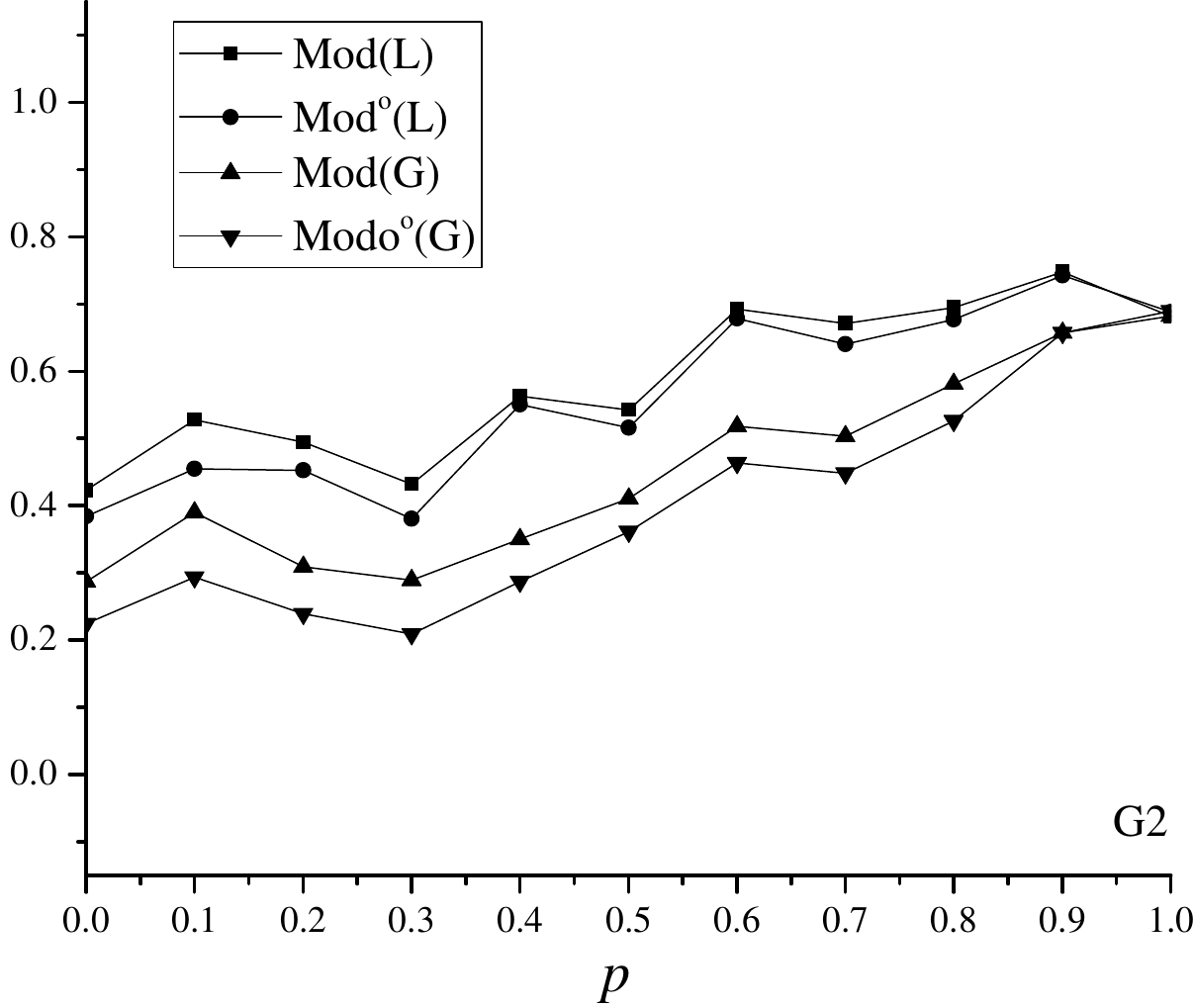}
      \hspace*{0.2cm}
      \includegraphics[width=0.30\textwidth]{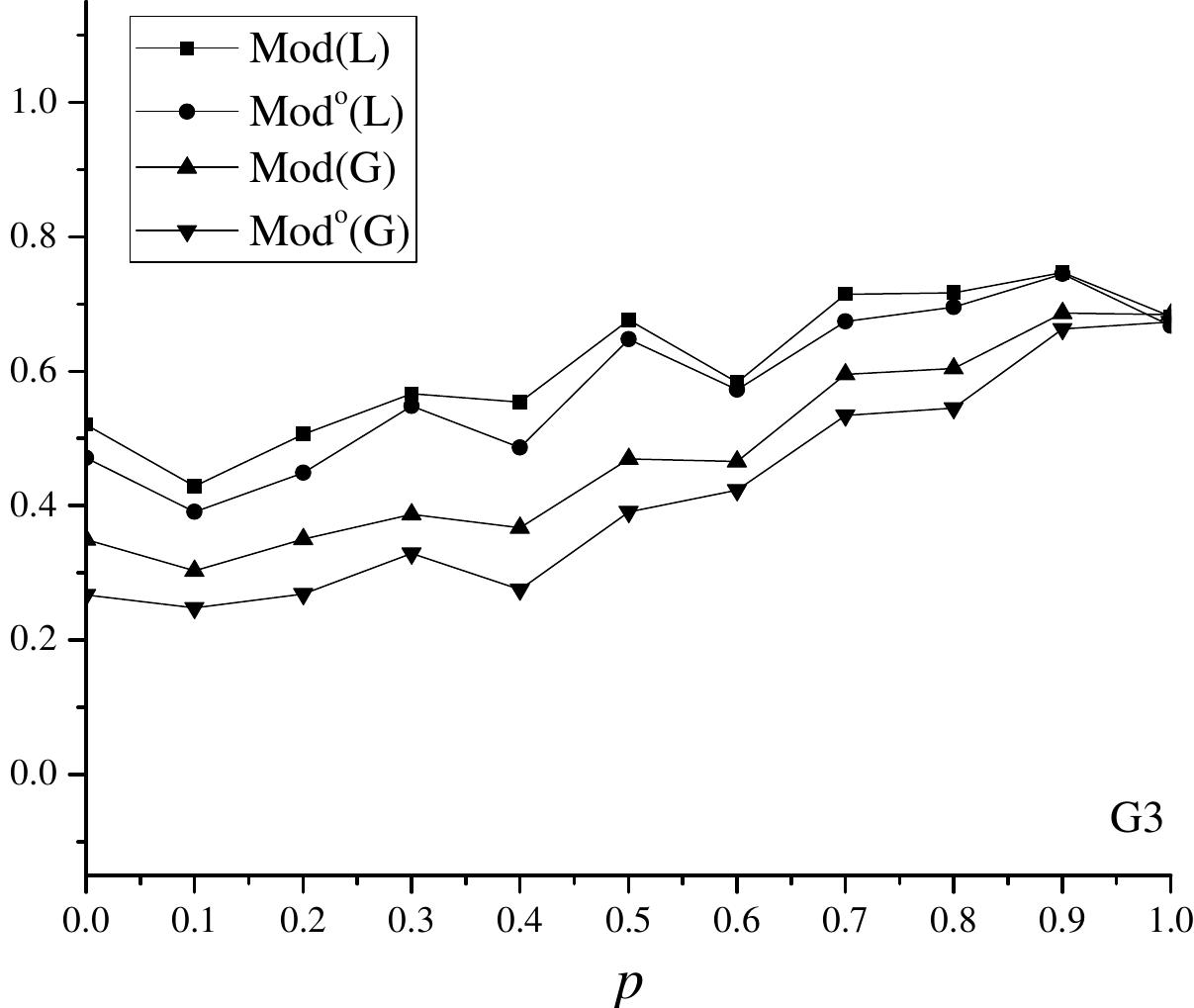} 
      \label{fig_p}
  }
  \end{minipage}
  \caption{
    The parameter $p \in (0, 1)$ controls the correlation of community structure across synthetic social networks. 
    As the value of $p$ steadily increases, the correlation between communities across these networks progressively intensifies.
    }
  \label{fig_p_syntheticexp}
\end{figure*}

To examine the impact of the correlation between social networks, we also conducted experiments wherein the parameter $p$ is progressively adjusted from 0.1 to 0.9 within the benchmark framework.
As shown in Fig.\ref{fig_p_syntheticexp} , it is evident that as the value of parameter $p$ gradually increases, there is a corresponding improvement in the experimental outcomes of various metrics.
This substantiates the assertion that the stronger the correlation amongst user communities across networks, 
the more advantageous it becomes for the integration of communities across networks, 
thereby engendering superior global community structure.

As mentioned above, we use NMI metric to measure consistency between the global community and each social network community.
The experimental results of NMI in the synthetic dataset are shown in Tab.\ref{tab_communityfusionnmiinsyntheticdata}.
The Tab.\ref{tab_communityfusionnmiinsyntheticdata} demonstrates that, through fusion in this study, the global community achieves a higher level of fusion compared to the other works.
Since the community detection in this paper focuses on fusing communities based on overlapping users, the similarity of the communities in each social network is primarily reflected in the consistency of the community structure, which hinges on overlapping users. 
Lastly, the table does not include a comparison with uniCDMN as it exhibits a fixed community structure shared across all social networks, leading to complete fusion. 
uniCDMN model cannot decrease the fusion degree within each social network to improve community quality.
\begin{table}[htb]
  \caption{Comparison of Community Fusion, the number of communities $K=20$}
  \label{tab_communityfusionnmiinsyntheticdata}
  \centering
  \resizebox{0.50\textwidth}{!}{%
  \begin{tabular}{c|cc|c|cc|c|cc}
  \toprule
  Network             & Method   & NMI            & Network             & Method   & NMI            & Network             & Method   & NMI            \\ \midrule
  \multirow{5}{*}{G1} & MultiNMF & 0.604          & \multirow{5}{*}{G2} & MultiNMF & 0.639          & \multirow{5}{*}{G3} & MultiNMF & 0.624          \\
                      & coNMF    & 0.721          &                     & coNMF    & 0.723          &                     & coNMF    & 0.760          \\
                      & coNMTF   & 0.804          &                     & coNMTF   & 0.754          &                     & coNMTF   & 0.754          \\
                      & MCD      & 0.622          &                     & MCD      & 0.692          &                     & MCD      & 0.659          \\
                      & HMCD     & \textbf{0.809} &                     & HMCD     & \textbf{0.801} &                     & HMCD     & \textbf{0.846} \\ 
  \bottomrule
  \end{tabular}%
  }
\end{table}

\subsection{Results Analysis on Real-World Dataset}
\subsubsection{Community Quality Evaluation}\label{section_communityquality}

\begin{table*}[!htbp]
  \centering 
  \caption{Comparison of Community Modularity and Compactness Based on Topology Attributes} 
  \label{tab_relation_com_exp}
  \resizebox{0.80\textwidth}{!}{%
  \begin{tabular}{ccccccc|ccccc}
  \toprule
  \multirow{2}{*}{Network}   & \multirow{2}{*}{Method} & \multicolumn{5}{c|}{$K=40$}                                                                                                     & \multicolumn{5}{c}{$K=80$}                                                                                              \\ \cmidrule(l){3-12} 
                             &                         & mod(L)                  & mod\textsuperscript{o}(L)& mod(G)                  & mod\textsuperscript{o}(G)& comp                  & mod(L)                  & mod\textsuperscript{o}(L)& mod(G)                  & mod\textsuperscript{o}(G)& comp      \\ \midrule
  \multirow{6}{*}{Twitter}   & MultiNMF                & 0.216 ±.010             & 0.238 ±.011              & 0.215 ±.009             & 0.245 ±.008              & 1437 ±098             & 0.209 ±.007             & 0.226 ±.013              & 0.178 ±.013             & 0.199 ±.017              & 1361 ±158 \\ 
                             & coNMF                   & 0.169 ±.004             & 0.182 ±.005              & 0.137 ±.009             & 0.162 ±.010              & 0851 ±055             & 0.168 ±.004             & 0.174 ±.005              & 0.128 ±.007             & 0.148 ±.009              & 1029 ±086  \\ 
                             & uniCDMN                 & 0.210 ±.018             & 0.232 ±.016              & 0.210 ±.018             & 0.232 ±.016              & 1334 ±085             & 0.198 ±.005             & 0.228 ±.004              & 0.198 ±.005             & 0.228 ±.004              & 1469 ±083  \\ 
                             & coNMTF                  & 0.223 ±.020             & 0.231 ±.025              & 0.179 ±.015             & 0.201 ±.015              & 1134 ±095             & 0.212 ±.014             & 0.221 ±.019              & 0.152 ±.007             & 0.177 ±.013              & 1078 ±071  \\ 
                             & MCD                     & 0.226 ±.004             & \textbf{0.286 ±.007}     & 0.163 ±.006             & 0.204 ±.008              & 0945 ±093             & 0.220 ±.006             & 0.254 ±.002              & 0.154 ±.004             & 0.196 ±.005              & 1022 ±075  \\ 
                             & HMCD                    & \textbf{0.249 ±.009}    & \underline{0.266 ±.010}  & \textbf{0.247 ±.010}    & \textbf{0.265 ±.010}     & \textbf{1517 ±093}    & \textbf{0.230 ±.003}    & \textbf{0.266 ±.005}     & \textbf{0.224 ±.002}    & \textbf{0.236 ±.004}     & \textbf{1560 ±075}  \\ \midrule
  \multirow{6}{*}{Instagram} & MultiNMF                & 0.299 ±.008             & 0.348 ±.012              & 0.267 ±.005             & 0.342 ±.005              & 0955 ±063             & 0.279 ±.005             & 0.328 ±.005              & 0.222 ±.012             & 0.294 ±.020              & 0933 ±066  \\ 
                             & coNMF                   & 0.307 ±.007             & 0.331 ±.009              & 0.232 ±.009             & 0.262 ±.009              & 0771 ±036             & 0.290 ±.004             & 0.320 ±.007              & 0.223 ±.009             & 0.249 ±.017              & 0977 ±024   \\ 
                             & uniCDMN                 & 0.279 ±.016             & 0.343 ±.017              & 0.279 ±.016             & 0.343 ±.017              & 0924 ±054             & 0.272 ±.017             & 0.348 ±.007              & 0.272 ±.017             & 0.348 ±.007              & 01083 ±077  \\ 
                             & coNMTF                  & 0.379 ±.007             & 0.414 ±.005              & 0.273 ±.013             & 0.301 ±.015              & 0882 ±010             & 0.370 ±.010             & 0.402 ±.010              & \textbf{0.295 ±.043}    & 0.310 ±.039              & 01045 ±058  \\ 
                             & MCD                     & \textbf{0.390 ±.006}    & \textbf{0.435 ±.006}     & 0.227 ±.012             & 0.324 ±.008              & 0605 ±064             & \textbf{0.375 ±.008}    & \textbf{0.413 ±.008}     & 0.225 ±.010             & 0.320 ±.006              & 0693 ±032  \\  
                             & HMCD                    & \underline{0.336 ±.010} & \underline{0.388 ±.014}  & \textbf{0.332 ±.010}    & \textbf{0.385 ±.014}     & \textbf{1100 ±047}    & \underline{0.298 ±.006} & \underline{0.357 ±.007}  & \underline{0.293 ±.006} & \textbf{0.355 ±.007}     & \textbf{1136 ±017}  \\ \midrule
  \multirow{6}{*}{Tumblr}    & MultiNMF                & \textbf{0.502 ±.002}    & 0.253 ±.021              & \textbf{0.415 ±.004}    & 0.288 ±.003              & \textbf{0042 ±004}    & \textbf{0.448 ±.005}    & 0.271 ±.080              & \textbf{0.381 ±.008}    & 0.301 ±.035              & \textbf{0044 ±002}     \\ 
                             & coNMF                   & 0.413 ±.012             & 0.302 ±.035              & \textbf{0.415 ±.021}    & 0.257 ±.041              & 0033 ±005             & 0.357 ±.014             & 0.283 ±.048              & 0.349 ±.011             & 0.230 ±.062              & 0024 ±005     \\ 
                             & uniCDMN                 & 0.307 ±.031             & 0.281 ±.102              & 0.307 ±.031             & 0.281 ±.102              & 0012 ±002             & 0.239 ±.004             & 0.314 ±.143              & 0.239 ±.004             & 0.314 ±.143              & 0015 ±005     \\ 
                             & coNMTF                  & 0.462 ±.007             & 0.311 ±.020              & 0.356 ±.021             & 0.301 ±.120              & 0026 ±003             & 0.411 ±.015             & 0.295 ±.043              & 0.328 ±.021             & 0.282 ±.079              & 0023 ±003     \\ 
                             & MCD                     & 0.385 ±.008             & \textbf{0.404 ±.040}     & 0.322 ±.009             & 0.257 ±.035              & 0024 ±002             & 0.335 ±.014             & \textbf{0.369 ±.021}     & 0.330 ±.010             & 0.230 ±.023              & 0030 ±001     \\ 
                             & HMCD                    & \underline{0.368 ±.036} & \underline{0.259 ±.092}  & \underline{0.367 ±.025} & \textbf{0.419 ±.072}     & \underline{0027 ±008} & \underline{0.371 ±.021} & \underline{0.303 ±.057}  & \underline{0.359 ±.024} & \textbf{0.319 ±.073}     & \underline{0026 ±005}     \\ 
  \bottomrule
  \end{tabular}%
  }
\end{table*}

\begin{table*}[!htbp]
  \centering 
  \caption{Comparison of Community Modularity and Density Based on Content Attributes}
  \label{tab_content_com_exp}
  \resizebox{0.80\textwidth}{!}{%
  \begin{tabular}{ccccccc|ccccc}
  \toprule
  \multirow{2}{*}{Network}   & \multirow{2}{*}{Method} & \multicolumn{5}{c|}{$K=40$}                                                                                                     & \multicolumn{5}{c}{$K=80$}                                                                                        \\ \cmidrule(l){3-12} 
                             &                         & mod(L)                  & mod\textsuperscript{o}(L)& mod(G)               & mod\textsuperscript{o}(G)& dens                 & mod(L)                  & mod\textsuperscript{o}(L)& mod(G)               & mod\textsuperscript{o}(G)& dens        \\ \midrule
  \multirow{6}{*}{Twitter}   & MultiNMF                & 0.269 ±.006             & 0.269 ±.009              & 0.257 ±.005          & 0.265 ±.007              & 0.249 ±.023          & 0.252 ±.010             & 0.251 ±.014              & 0.210 ±.011          & 0.212 ±.014              & 0.247 ±.010 \\ 
                             & coNMF                   & 0.273 ±.013             & 0.217 ±.013              & 0.163 ±.005          & 0.175 ±.009              & 0.203 ±.004          & 0.240 ±.008             & 0.198 ±.008              & 0.137 ±.007          & 0.155 ±.009              & 0.200 ±.006 \\ 
                             & uniCDMN                 & 0.229 ±.017             & 0.233 ±.017              & 0.229 ±.017          & 0.233 ±.017              & 0.194 ±.059          & 0.211 ±.007             & 0.227 ±.004              & 0.211 ±.007          & 0.227 ±.004              & 0.198 ±.003 \\ 
                             & coNMTF                  & \textbf{0.387 ±.013}    & \textbf{0.308 ±.005}     & 0.235 ±.020          & 0.218 ±.022              & 0.203 ±.005          & \textbf{0.381 ±.028}    & \textbf{0.311 ±.010}     & 0.180 ±.023          & 0.184 ±.013              & 0.196 ±.003 \\ 
                             & MCD                     & 0.149 ±.006             & 0.210 ±.006              & 0.171 ±.007          & 0.215 ±.011              & 0.207 ±.008          & 0.145 ±.005             & 0.196 ±.011              & 0.167 ±.003          & 0.213 ±.004              & 0.202 ±.006 \\ 
                             & HMCD                    & \underline{0.295 ±.008} & \underline{0.284 ±.009}  & \textbf{0.292 ±.008} & \textbf{0.282 ±.008}     & \textbf{0.267 ±.016} & \underline{0.263 ±.003} & \underline{0.256 ±.004}  & \textbf{0.258 ±.003} & \textbf{0.252 ±.003}     & \textbf{0.261 ±.006} \\ \midrule
  \multirow{6}{*}{Instagram} & MultiNMF                & 0.303 ±.005             & 0.379 ±.010              & 0.271 ±.003          & 0.349 ±.006              & 0.207 ±.009          & 0.279 ±.003             & 0.353 ±.007              & 0.226 ±.006          & 0.307 ±.017              & 0.214 ±.014 \\ 
                             & coNMF                   & 0.341 ±.005             & 0.351 ±.020              & 0.200 ±.006          & 0.274 ±.009              & 0.183 ±.005          & 0.316 ±.006             & 0.358 ±.088              & 0.179 ±.006          & 0.257 ±.045              & 0.181 ±.006 \\ 
                             & uniCDMN                 & 0.273 ±.009             & 0.343 ±.019              & 0.273 ±.009          & 0.338 ±.019              & 0.190 ±.008          & 0.255 ±.013             & 0.340 ±.008              & 0.255 ±.013          & 0.340 ±.008              & 0.191 ±.005 \\ 
                             & coNMTF                  & 0.306 ±.078             & 0.338 ±.052              & 0.245 ±.006          & 0.313 ±.020              & 0.180 ±.005          & \textbf{0.386 ±.058}    & \textbf{0.443 ±.029}     & 0.256 ±.045          & 0.314 ±.043              & 0.149 ±.003 \\ 
                             & MCD                     & 0.279 ±.003             & 0.373 ±.002              & 0.240 ±.009          & 0.343 ±.012              & 0.185 ±.001          & 0.268 ±.004             & 0.373 ±.006              & 0.218 ±.004          & 0.340 ±.005              & 0.182 ±.003 \\ 
                             & HMCD                    & \textbf{0.346 ±.006}    & \textbf{0.410 ±.013}     & \textbf{0.340 ±.007} & \textbf{0.402 ±.128}     & \textbf{0.214 ±.010} & \underline{0.298 ±.003} & \underline{0.380 ±.007}  & \textbf{0.292 ±.004} & \textbf{0.374 ±.006}     & \textbf{0.214 ±.001} \\ \midrule
  \multirow{6}{*}{Tumblr}    & MultiNMF                & 0.131 ±.012             & \textbf{0.521 ±.080}     & 0.089 ±.006          & 0.214 ±.032              & 0.229 ±.018          & 0.118 ±.006             & \textbf{0.451 ±.059}     & 0.068 ±.003          & 0.241 ±.062              & 0.220 ±.008 \\ 
                             & coNMF                   & 0.179 ±.012             & 0.454 ±.035              & 0.097 ±.021          & 0.204 ±.041              & 0.184 ±.003          & 0.107 ±.014             & 0.404 ±.048              & 0.066 ±.011          & 0.222 ±.062              & 0.200 ±.003 \\ 
                             & uniCDMN                 & 0.127 ±.011             & 0.244 ±.129              & 0.127 ±.011          & 0.244 ±.129              & 0.222 ±.024          & 0.092 ±.007             & 0.347 ±.197              & 0.092 ±.007          & 0.347 ±.197              & \textbf{0.269 ±.007} \\ 
                             & coNMTF                  & 0.206 ±.020             & 0.464 ±.010              & 0.096 ±.009          & 0.297 ±.148              & 0.198 ±.013          & \textbf{0.212 ±.004}    & 0.439 ±.024              & 0.067 ±.005          & 0.261 ±.102              & 0.196 ±.029 \\ 
                             & MCD                     & 0.057 ±.005             & 0.210 ±.002              & 0.072 ±.004          & 0.399 ±.049              & \textbf{0.237 ±.011} & 0.054 ±.003             & 0.417 ±.014              & 0.066 ±.004          & \textbf{0.470 ±.014}     & 0.216 ±.002 \\ 
                             & HMCD                    & \textbf{0.221 ±.011}    & \underline{0.375 ±.043}  & \textbf{0.215 ±.012} & \textbf{0.400 ±.092}     & \textbf{0.237 ±.027} & \underline{0.111 ±.009} & \underline{0.269 ±.151}  & \textbf{0.096 ±.008} & \underline{0.274 ±.104}  & \underline{0.252 ±.023}\\
  \bottomrule
  \end{tabular}%
  }
\end{table*}

\begin{table*}[htbp]
  \centering
  \caption{Comparison of Community Quality Only on Twitter and Instagram}
  \label{tab_relation_content_com_exp2}
  \resizebox{0.93\textwidth}{!}{
  \begin{tabular}{ccccc|ccc|ccc|ccc}
  \toprule
  \multirow{3}{*}{Network}   & \multirow{3}{*}{Method} & \multicolumn{6}{c|}{Topology}                                                                                                                      & \multicolumn{6}{c}{Content}                                                                                                          \\ \cmidrule(l){3-14} 
                             &                         & \multicolumn{3}{c|}{$K=40$}                                             & \multicolumn{3}{c|}{$K=80$}                                              & \multicolumn{3}{c|}{$K=40$}                                          & \multicolumn{3}{c}{$K=80$}                                             \\ \cmidrule(l){3-14} 
                             &                         & mod(G)                & mod\textsuperscript{o}(G) & comp                & mod(G)               & mod\textsuperscript{o}(G) & comp                  & mod(G)              & mod\textsuperscript{o}(G) & dens                & mod(G)              & mod\textsuperscript{o}(G) & dens                \\ \midrule
  \multirow{6}{*}{Twitter}   & MultiNMF                & 0.20 ±.010            & 0.23 ±.015                & 1271 ±086           & 0.18 ±.028           & 0.20 ±.031                & 1258 ±084             & 0.25 ±.004          & 0.26 ±.007                & 0.25 ±.023          & 0.21 ±.032          & 0.21 ±.037                & 0.23 ±.016          \\
                             & coNMF                   & 0.14 ±.008            & 0.16 ±.008                & 0850 ±081           & 0.13 ±.011           & 0.15 ±.014                & 1053 ±091             & 0.17 ±.004          & 0.18 ±.006                & 0.23 ±.044          & 0.14 ±.009          & 0.16 ±.012                & 0.20 ±.006          \\
                             & uniCDMN                 & 0.21 ±.003            & 0.23 ±.006                & 1310 ±042           & 0.20 ±.013           & 0.23 ±.009                & 1494 ±069             & 0.23 ±.005          & 0.24 ±.006                & 0.20 ±.002          & 0.22 ±.020          & 0.23 ±.011                & 0.19 ±.004          \\
                             & coNMTF                  & 0.17 ±.009            & 0.20 ±.009                & 1208 ±210           & 0.16 ±.005           & 0.19 ±.009                & 1092 ±082             & 0.22 ±.024          & 0.21 ±.009                & 0.20 ±.009          & 0.19 ±.018          & 0.19 ±.009                & 0.20 ±.005          \\
                             & MCD                     & 0.15 ±.008            & 0.19 ±.012                & 1017 ±120           & 0.13 ±.009           & 0.16 ±.019                & 0919 ±089             & 0.19 ±.029          & 0.20 ±.031                & 0.20 ±.015          & 0.17 ±.008          & 0.17 ±.009                & 0.20 ±.005          \\
                             & HMCD                    & \textbf{0.24 ±.008}   & \textbf{0.26 ±.010}       & \textbf{1503 ±072}  & \textbf{0.22 ±.008}  & \textbf{0.23 ±.007}       & \textbf{1555 ±071}     & \textbf{0.29 ±.004} & \textbf{0.28 ±.005}       & \textbf{0.27 ±.019} & \textbf{0.26 ±.004} & \textbf{0.25 ±.008}       & \textbf{0.25 ±.003} \\ \midrule
  \multirow{6}{*}{Instagram} & MultiNMF                & 0.27 ±.028            & 0.34 ±.025                & 0884 ±093           & 0.23 ±.020           & 0.29 ±.018                & 0965 ±083             & 0.27 ±.017          & 0.36 ±.019                & 0.19 ±.018          & 0.23 ±.010          & 0.30 ±.019                & 0.20 ±.010          \\
                             & coNMF                   & 0.24 ±.017            & 0.28 ±.015                & 0826 ±076           & 0.24 ±.011           & 0.27 ±.026                & 1047 ±050             & 0.21 ±.014          & 0.29 ±.010                & 0.18 ±.005          & 0.19 ±.010          & 0.27 ±.024                & 0.18 ±.004          \\
                             & uniCDMN                 & 0.28 ±.007            & 0.34 ±.007                & 0908 ±028           & 0.27 ±.018           & 0.34 ±.014                & 1138 ±071             & 0.28 ±.009          & 0.34 ±.007                & 0.19 ±.005          & 0.26 ±.012          & 0.33 ±.014                & 0.18 ±.003          \\
                             & coNMTF                  & 0.28 ±.010            & 0.31 ±.009                & 0931 ±065           & 0.28 ±.018           & 0.31 ±.016                & 1091 ±030             & 0.24 ±.043          & 0.32 ±.007                & 0.18 ±.004          & 0.24 ±.016          & 0.31 ±.017                & 0.18 ±.002          \\
                             & MCD                     & 0.22 ±.010            & 0.28 ±.020                & 0810 ±178           & 0.20 ±.011           & 0.26 ±.008                & 0769 ±082             & 0.31 ±.008          & 0.29 ±.053                & 0.18 ±.011          & 0.22 ±.014          & 0.25 ±.009                & 0.19 ±.007          \\
                             & HMCD                    & \textbf{0.34 ±.013}   & \textbf{0.38 ±.038}       & \textbf{1094 ±086}   & \textbf{0.30 ±.006}  & \textbf{0.36 ±.008}       & \textbf{1158 ±032}    & \textbf{0.34 ±.008} & \textbf{0.41 ±.014}       & \textbf{0.21 ±.007} & \textbf{0.32 ±.047} & \textbf{0.38 ±.008}       & \textbf{0.22 ±.005} \\ \bottomrule
  \end{tabular}
  }
\end{table*}

\begin{figure*}[htb]
  \centering
  \begin{minipage}[b]{0.82\textwidth}
    \centering
    \subfloat[modularity]{
      \includegraphics[width=0.28\textwidth]{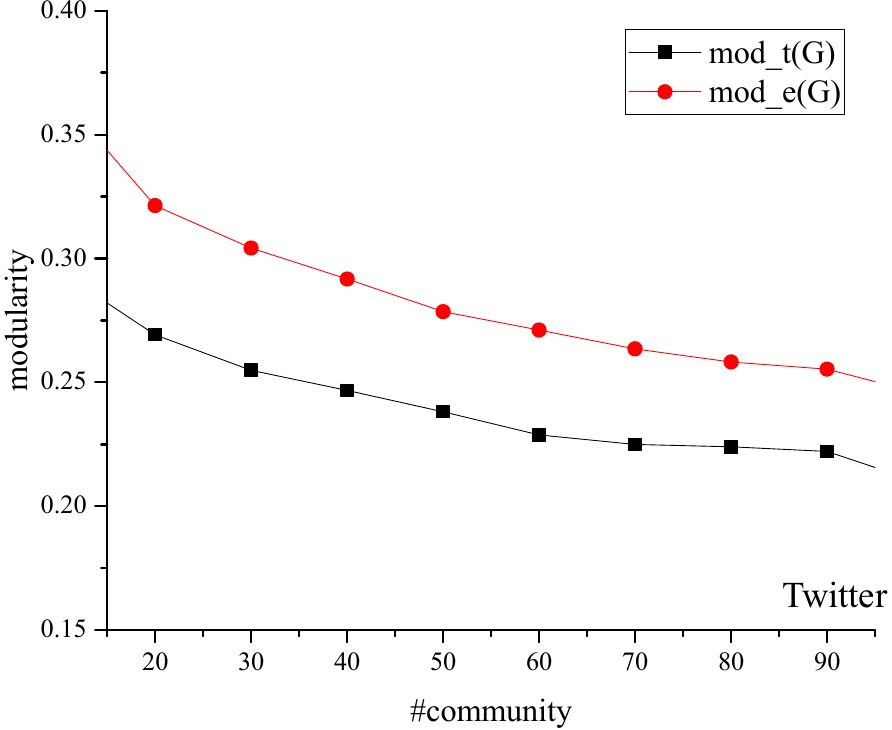}
      \hspace*{0.8cm}
      \includegraphics[width=0.28\textwidth]{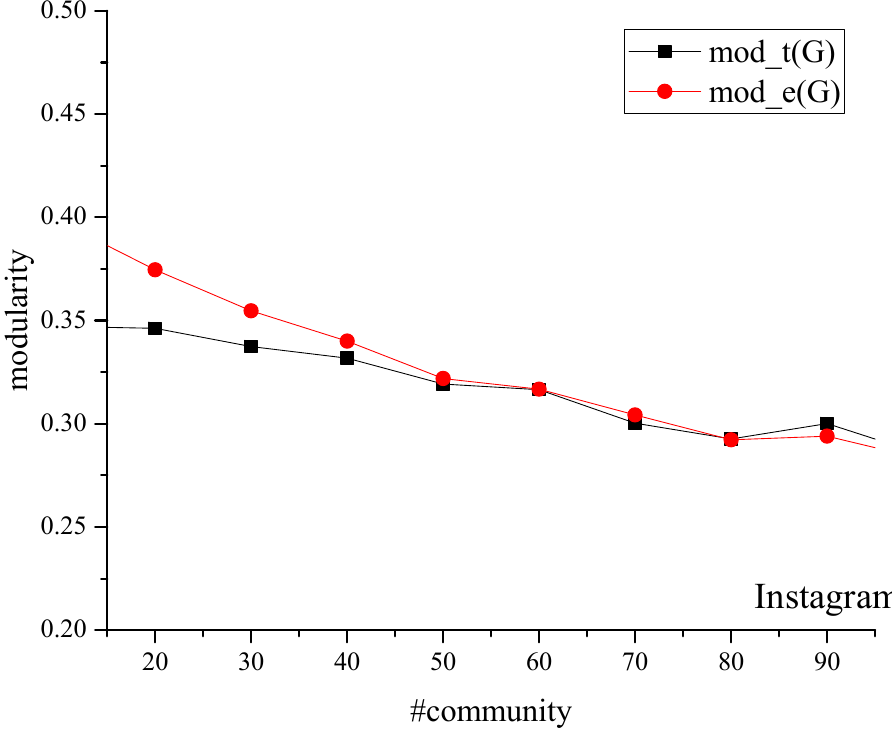}
      \hspace*{0.8cm}
      \includegraphics[width=0.28\textwidth]{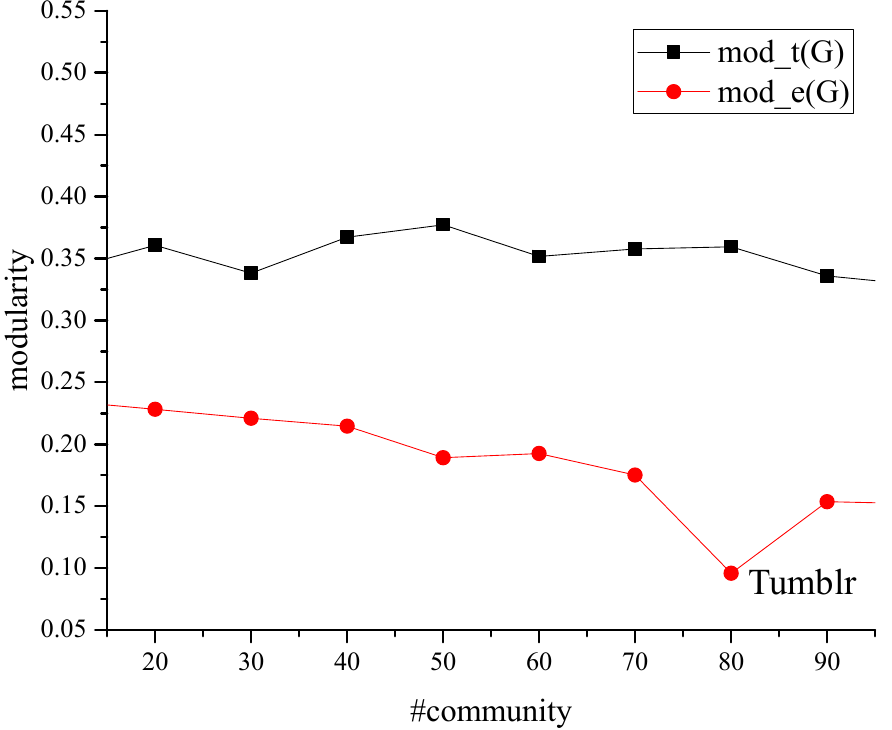} 
      \label{fig_numberofcommunityexp_mod}
  }
  \end{minipage}
  \begin{minipage}[b]{0.82\textwidth}
    \centering
    \subfloat[compactness]{
      \includegraphics[width=0.28\textwidth]{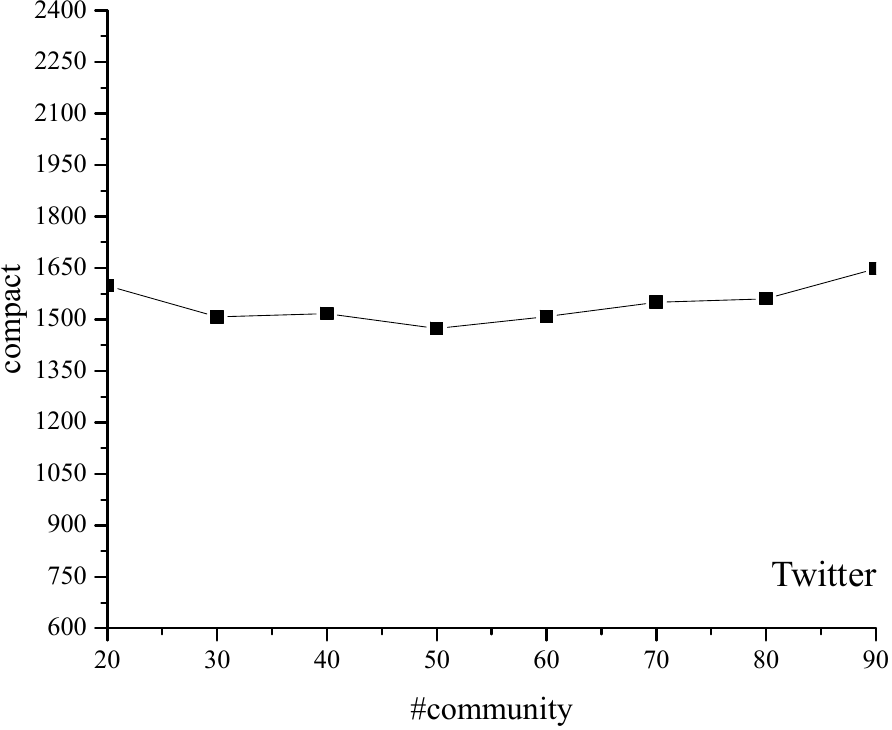}
      \hspace*{0.8cm}
      \includegraphics[width=0.28\textwidth]{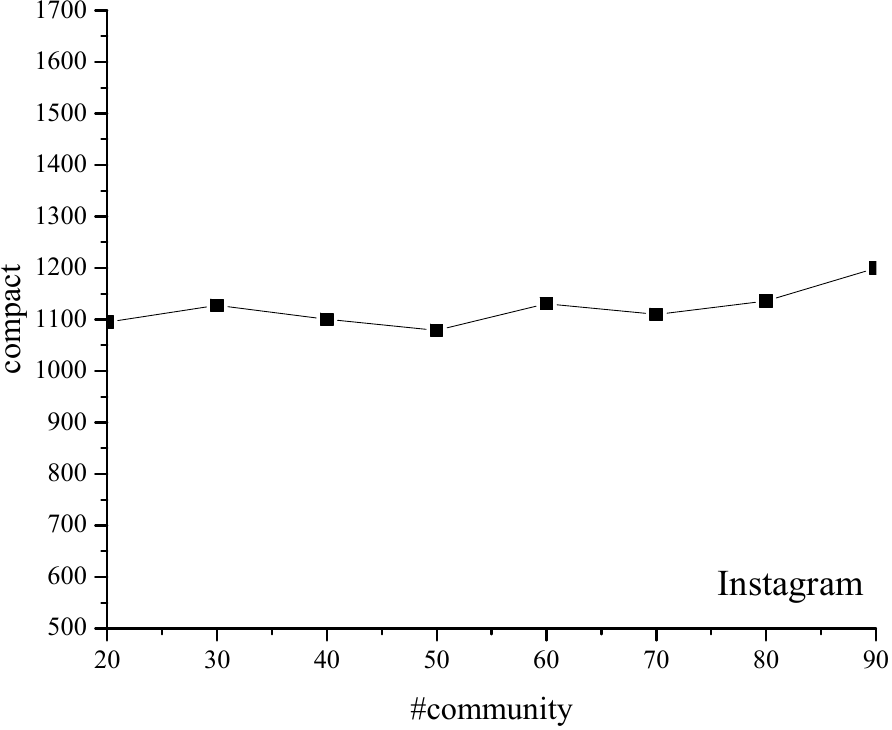}
      \hspace*{0.8cm}
      \includegraphics[width=0.28\textwidth]{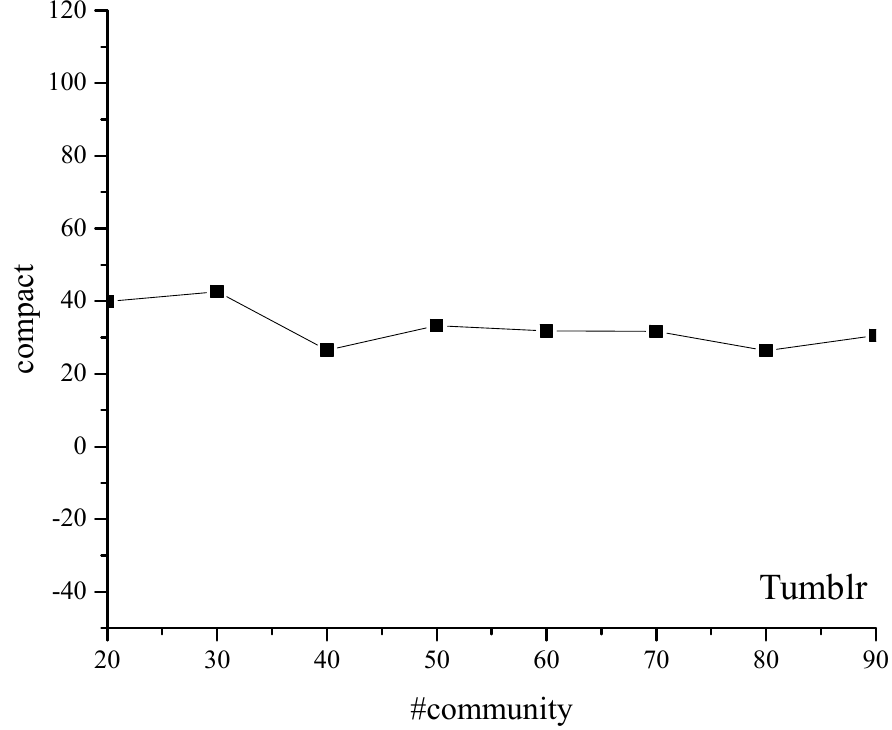} 
      \label{fig_numberofcommunityexp_comp}
  }
  \end{minipage}
  \begin{minipage}[b]{0.82\textwidth}
    \centering
    \subfloat[density]{
      \includegraphics[width=0.28\textwidth]{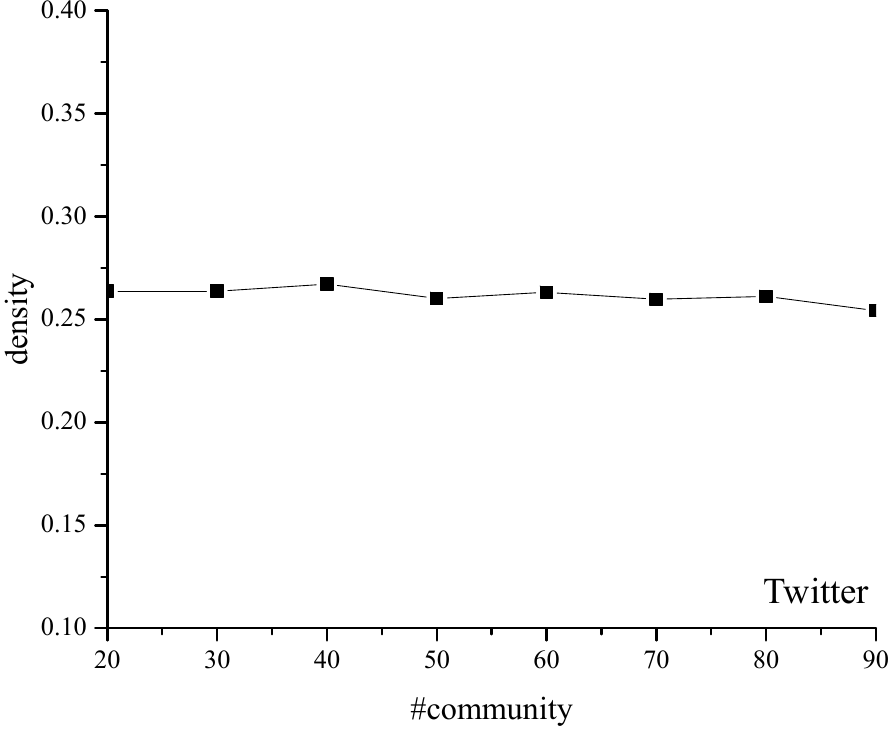}
      \hspace*{0.8cm}
      \includegraphics[width=0.28\textwidth]{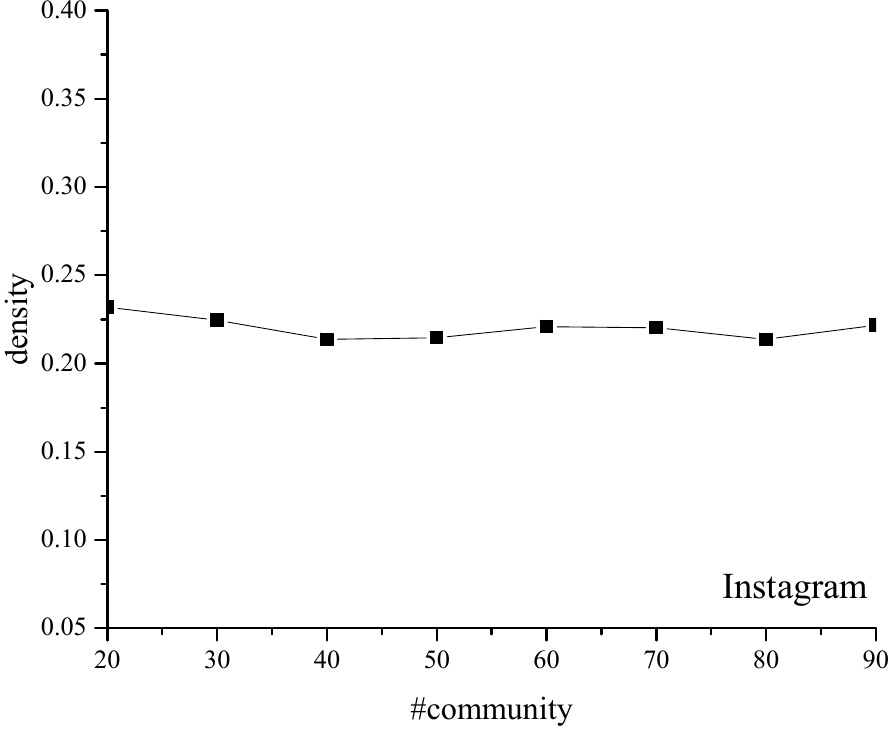}
      \hspace*{0.8cm}
      \includegraphics[width=0.28\textwidth]{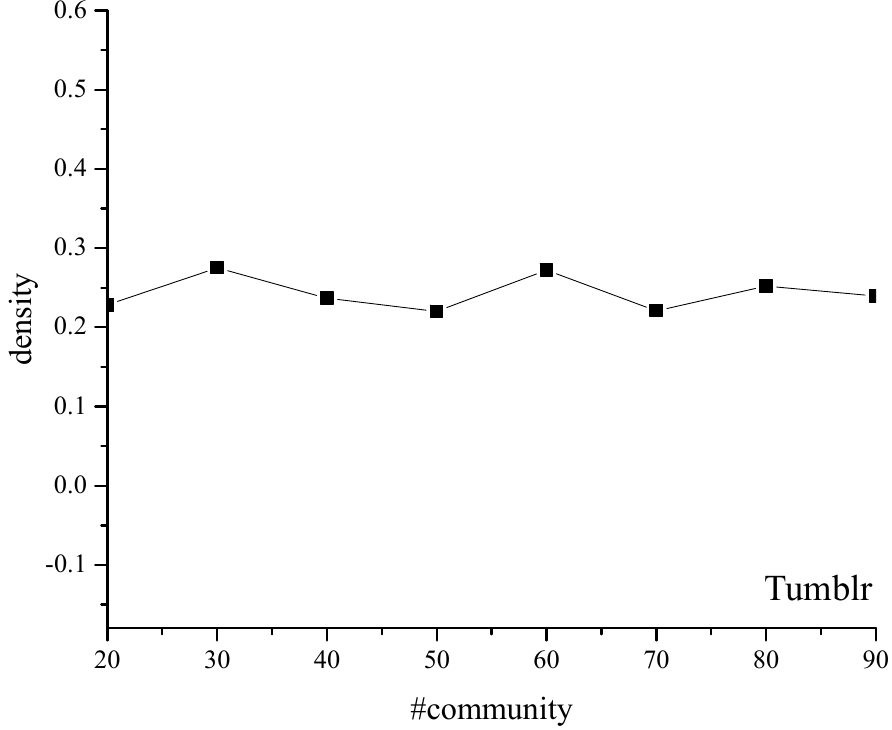} 
      \label{fig_numberofcommunityexp_dens}
  }
  \end{minipage}
  \caption{Community quality with different $K$ in community detection. The modularity of the global community based on topology and content attributes is denoted
  as mod\_t(G) and mod\_e(G).}
  \label{fig_numberofcommunityexp}
\end{figure*}

This section evaluates the community quality based on its topology and content attributes. 

The experiment involves four types of modularity, as described in the previous section: \textbf{mod(L)}, \textbf{mod(G)}, \textbf{mod\textsuperscript{o}(L)} and \textbf{mod\textsuperscript{o}(G)}. 

In the previous section, we also established evaluation metrics for both topology and content attributes.
We apply these metrics, which are specific to the global community structure across social networks, in our experiments.
The metrics are denoted as ``\textbf{comp}'' for compactness and ``\textbf{dens}'' for density.

First, this experiment assesses the community's quality by examining its content attributes. 
Based on the user size in the dataset, and referring to the work\cite{philip2015mcd}, the study selects the number of community divisions from a range of 10 to 100. 
This paper specifically chooses 40 and 80, near the two ends of this range, as the numbers for community divisions to be compared. 
The effects of other division numbers on this paper's model are presented in the experiment. 
Tab.\ref{tab_relation_com_exp} shows the experimental results when the number of community divisions is either 40 or 80.

Overall, the quality of the global community over topology attributes identified by the HMCD model in this paper surpasses that of other models. 
However, this model does not hold an absolute advantage in every aspect of community quality, specifically mod(L) and mod\textsuperscript{o}(L), across all networks. 
This situation could be tied to the model's design, which aims for consistency in community structure across all social networks. 
By emphasizing consistency, the model may overlook unique features and nuances that distinguish each community within their specific social networks. 
Consequently, the global community obtained might not fully encapsulate the essence of the original community structure in each social network. 
In particular, the model's modularity based on overlapping users, mod\textsuperscript{o}(G), compactness, and density, as shown in the table, significantly outperform other models. 
Since a community is essentially portrayed as a user aggregation structure, the HMCD is designed with overlapping users in mind. 
It tends to reveal the aggregated structure between overlapping users across social networks. 
Likewise, improved results are also achieved on Twitter and Instagram in terms of compactness for overlapping users. 
Regrettably, Tumblr has not delivered the same positive results as other social networks. 
This could be attributed to the scarcity of intersecting users within our Tumblr dataset, leading to a conspicuous absence of substantial communal linkages among users. 
The process of community detection for overlapping users will inevitably influence community quality.

We also evaluate the community quality based on the community's content attributes. 
The Fig.\ref{tab_content_com_exp} displays the experimental results when the number of communities is segmented into 40 and 80.
The global community identified by the HMCD model in this study significantly outperforms other models.
However, the model does not always achieve an absolute advantage in every single network.
At the same time, due to the sparse number of overlapping users on Tumblr, the experimental results on the Tumblr network also decrease.

From the aforementioned experimental analysis, it is discernible that the scarcity of Tumblr data has exerted a negative impact on the results. 
To more effectively demonstrate the validity of the model presented, we have chosen to exclude Tumblr data, conducting our experiment solely on the Twitter and Instagram datasets.
As illustrated in Table.\ref{tab_relation_content_com_exp2}, our model demonstrates the superior performance over others, particularly with respect to global community metrics, namely mod(G), mod(G)\textsuperscript{o}, dens, and comp.
\subsubsection{The Number of Communities}
To examine the impact of the number of communities $K$ on results, we conducted experiments wherein we manipulated the number of communities detected, as shown in Fig.\ref{fig_numberofcommunityexp}.
The modularity of the global community based on topology and content attributes is denoted as \textbf{mod\_t(G)} and \textbf{mod\_e(G)}.
It can be found from Fig.\ref{fig_numberofcommunityexp} that the metrics used in the experiment remained relatively stable despite a change in the number of communities. 
This occurrence is likely attributed to the sparse distribution of user data in social networks.
For instance, with regard to the topology attribute investigated in this paper, the connection relationships among some users exhibit a significant degree of concentration.
These users easily aggregate with connected neighbors during the process of community detection, subsequently constituting a community. 
The community structures formed by these users in large numbers are relatively stable compared with the remaining users, resulting in less noticeable changes in the overall community structure of the network.
\subsubsection{Community Fusion Evaluation}
\begin{table}[htb]
  \caption{Comparison of Community Fusion}
  \label{tab_communityfusionnmi}
  \centering
  \resizebox{0.34\textwidth}{!}{
  \begin{tabular}{cccc|cc}
  \toprule
  \multirow{2}{*}{Network} & \multirow{2}{*}{Method} & \multicolumn{2}{c|}{40} & \multicolumn{2}{c}{80} \\ \cline{3-6} 
                                  &                         & NMI\_t     & NMI\_e     & NMI\_t     & NMI\_e    \\ \midrule
  \multirow{5}{*}{Twitter}        & MultiNMF                & 0.632      & 0.660      & 0.644      & 0.660     \\
                                  & coNMF                   & 0.525      & 0.556      & 0.640      & 0.660     \\
                                  & coNMTF                  & 0.542      & 0.424      & 0.628      & 0.410     \\
                                  & MCD                     & 0.547      & 0.593      & 0.588      & 0.629     \\
                                  & HMCD                    & \textbf{0.959}      & \textbf{0.943}      & \textbf{0.958}      & \textbf{0.951}     \\ \midrule
  \multirow{5}{*}{Instagram}      & MultiNMF                & 0.640      & 0.677      & 0.632      & 0.641     \\
                                  & coNMF                   & 0.629      & 0.599      & 0.723      & 0.684     \\
                                  & coNMTF                  & 0.578      & 0.433      & 0.660      & 0.500     \\
                                  & MCD                     & 0.504      & 0.527      & 0.564      & 0.571     \\
                                  & HMCD                    & \textbf{0.957}      & \textbf{0.944}      & \textbf{0.953}      & \textbf{0.946}     \\ \midrule
  \multirow{5}{*}{Tumblr}         & MultiNMF                & 0.814      & 0.647      & 0.905      & 0.760     \\
                                  & coNMF                   & 0.863      & 0.839      & 0.927      & 0.908     \\
                                  & coNMTF                  & 0.744      & 0.690      & 0.823      & 0.765     \\
                                  & MCD                     & 0.751      & 0.805      & 0.807      & 0.848     \\
                                  & HMCD                    & \textbf{0.941}      & \textbf{0.918}      & \textbf{0.946}      & \textbf{0.929}     \\ \bottomrule
  \end{tabular}%
  }
\end{table}
The consistency between the global community and each social network community can reflect the model's ability to fuse communities across networks. 
In the experiment, we denote the metrics of community consistency over the topology and content attributes as \textbf{NMI\_t} and \textbf{NMI\_e} respectively. 
When the number of communities is 40 and 80, the consistency metrics between the global community and intrinsic community in each social network are displayed in Tab.\ref{tab_communityfusionnmi}.
While endeavoring to attain a heightened level of community fusion, our model ensures that the quality of the community remains at a comparatively superior performance.

\section{Conclusion}
This paper introduces a community detection model that is applicable to multiple social networks. 
The paper utilizes both the topology and content information of users in social networks to model their adjacency relations. 
A user identity alignment matrix is designed for the overlapping users in each social network, which more effectively integrates the community structures in different social networks and obtains high-quality global community structures for multiple social networks. 
Additionally, the paper also proves the convergence of the model parameters. 
Experiments verify the validity of the method proposed in this paper. 
The community detection method proposed in this paper provides a basis for the information diffusion across social networks and the analysis of users' behavior patterns for multiple social networks. 
The model presented in our paper revolves around the core of overlapping users for community detection, wherein the number or attribute relationships of these overlapping users' communities align with those in the regular user community detection. 
In the future, we aim to perpetuate the exploration of diversified forms of community detection, further accentuating the significance of overlapping users, and seeking suitable application scenarios. 
For instance, we consider overlapping users as anomalous users, who belong to a different number of communities compared to regular users, for the purpose of studying issues related to information security.




\bibliographystyle{IEEEtran}
\bibliography{IEEEabrv,referencepapers/mybibfile}

\newpage

\section*{Appendix}\label{section_communitydecomposition_proof}

We take the iterative update rules of parameters $\boldsymbol{{\rm X}}_k^s$ and $\boldsymbol{{\rm D}}_s^k$ as examples to demonstrate the convergence of the algorithm in this paper.

To prove that the updated iterative Eq.\ref{eq_xsolver} or Eq.\ref{eq_dsolver} makes the objective function Eq.\ref{eq_lagrangeglobalcommunitydecomposition} converge,
first of all, from the literature\cite{wang2011community}, the following conclusions can be drawn:

Theorem 1: 

Matrices $\boldsymbol{{\rm A}}$, $\boldsymbol{{\rm X}}$, $\boldsymbol{{\rm X}}'$ are non-negative, we have $-tr(\boldsymbol{{\rm X}}^T\boldsymbol{{\rm X}}) \leq 
-tr(\boldsymbol{{\rm A}}^{\rm T}\boldsymbol{{\rm Z}}) -tr(\boldsymbol{{\rm A}}^{\rm T}\boldsymbol{{\rm X}}')$, where $\boldsymbol{{\rm A}}_{i,j}=\boldsymbol{{\rm X}}_{i,j}'\ln\boldsymbol{{\rm X}}_{i,j}/\boldsymbol{{\rm X}}'_{i,j}$.

Theorem 2:  

Matrices $\boldsymbol{{\rm A}}$, $\boldsymbol{{\rm B}}$, $\boldsymbol{{\rm X}}$ are non-negative, we have  $-tr(\boldsymbol{{\rm B}}\boldsymbol{{\rm A}}^{\rm T}\boldsymbol{{\rm A}}\boldsymbol{{\rm X}}) \leq 
-tr(\boldsymbol{{\rm B}}\boldsymbol{{\rm X}}'^{\rm T}\boldsymbol{{\rm A}}\boldsymbol{{\rm Z}})-tr(\boldsymbol{{\rm B}}\boldsymbol{{\rm Z}}^{\rm T}\boldsymbol{{\rm A}}\boldsymbol{{\rm X}}')$, where $\boldsymbol{{\rm Z}}_{i,j}=\boldsymbol{{\rm X}}'_{i,j}\ln \boldsymbol{{\rm X}}_{i,j}/\boldsymbol{{\rm X}}'_{i,j} $.

Theorem 3:  

Matrices $\boldsymbol{{\rm A}}$, $\boldsymbol{{\rm B}}$, $\boldsymbol{{\rm X}}$ are non-negative, we have  $tr(\boldsymbol{{\rm B}}\boldsymbol{{\rm X}}'^{\rm T}\boldsymbol{{\rm A}}\boldsymbol{{\rm X}}) \leq 
\frac{1}{2}(tr(\boldsymbol{{\rm B}}\boldsymbol{{\rm Y}}^{\rm T}\boldsymbol{{\rm A}}\boldsymbol{{\rm X}}')+tr(\boldsymbol{{\rm B}}\boldsymbol{{\rm X}}'^{\rm T}\boldsymbol{{\rm A}}\boldsymbol{{\rm Y}})) $, where $\boldsymbol{{\rm Y}}_{i,j}=\boldsymbol{{\rm X}}_{i,j}^2/\boldsymbol{{\rm X}}'_{i,j}$.

Theorem 4: 

Matrices $\boldsymbol{{\rm A}}$, $\boldsymbol{{\rm X}}$, $\boldsymbol{{\rm X}}'$ are non-negative, we have $tr(\boldsymbol{{\rm P}}  \boldsymbol{{\rm A}}) \leq
\frac{1}{2}(tr(\boldsymbol{{\rm R}}\boldsymbol{{\rm A}}\boldsymbol{{\rm X}}'^{\rm T})) $, where $\boldsymbol{{\rm P}}_{k,j}=(\boldsymbol{{\rm X}}^{\rm T}\boldsymbol{{\rm X}})^2_{k,l}/(\boldsymbol{{\rm X}}'^{\rm T}\boldsymbol{{\rm X}}')_{k,l}$,
$\boldsymbol{{\rm R}}_{k,j}=(\boldsymbol{{\rm X}})_{k,l}^4/(\boldsymbol{{\rm X}}_{k,l}^3)$.

Theorem 5: 

Matrices $\boldsymbol{{\rm A}}$, $\boldsymbol{{\rm X}}$, $\boldsymbol{{\rm X}}'$ are non-negative, we have $tr(\boldsymbol{{\rm A}}\boldsymbol{{\rm X}}\boldsymbol{{\rm X}}^{\rm T}) \leq 
\frac{1}{2}(tr(\boldsymbol{{\rm R}}\boldsymbol{{\rm A}}\boldsymbol{{\rm X}'^{\rm T}})+tr(\boldsymbol{{\rm A}}\boldsymbol{{\rm X}}'\boldsymbol{{\rm X}}'^{\rm T}))$, where $\boldsymbol{{\rm R}}_{k,j}=(\boldsymbol{{\rm X}})_{k,l}^4/(\boldsymbol{{\rm X}}')_{k,l}^3$.

Theorem 6: 

When $x>0$, then $-x \leq -1-\ln x$ .

Theorem 7: 

When $x_i, y_i>0$, then $(\sum_i x_iy_i)^2 \leq (\sum_i x_i) (\sum_i x_iy_i^2)$.

Proof: 

1) After normalizing $\boldsymbol{{\rm X}}_k^s \leftarrow \boldsymbol{{\rm X}}_k^s\boldsymbol{{\rm Q}}_k^s$,\quad
$\boldsymbol{{\rm D}}_k^s \leftarrow (\boldsymbol{{\rm Q}}_k^s)^{-1}  \boldsymbol{{\rm D}}_k^s\\(\boldsymbol{{\rm Q}}_k^s)^{-1}$, the Lagrangian function of $\boldsymbol{{\rm X}}_k^s$ is obtained according to Eq.\ref{eq_lagrangeglobalcommunitydecomposition}:
\begin{equation}\label{eq_xglobalcommunitydecomposition_}
  \begin{aligned}
    &L(\boldsymbol{{\rm{X}}}_k^s)\\
    &=  \alpha_k^s tr(-\boldsymbol{{\rm{M}}}_k^s\boldsymbol{{\rm{X}}}_k^s{\boldsymbol{{\rm{D}}}_k^s}^{\rm T}{\boldsymbol{{\rm{X}}}_k^s}^{\rm T}
    -\boldsymbol{{\rm{X}}}_k^s\boldsymbol{{\rm{D}}}_k^s{\boldsymbol{{\rm{X}}}_k^s}^{\rm T}{\boldsymbol{{\rm{M}}}_k^s}^{\rm T}\\
    &\quad \quad \quad  +\boldsymbol{{\rm{X}}}_k^s\boldsymbol{{\rm{D}}}_k^s{\boldsymbol{{\rm{X}}}_k^s}^{\rm T}\boldsymbol{{\rm{X}}}_k^s{\boldsymbol{{\rm{D}}}_k^s}^{\rm T}{\boldsymbol{{\rm{X}}}_k^s}^{\rm T})\\
    &+ \gamma_k^s tr(-\boldsymbol{{\rm{H}}}_k^s\boldsymbol{{\rm{X}}}_k^s{\boldsymbol{{\rm{X}}}_k^s}^{\rm T}{\boldsymbol{{\rm{H}}}_k^s}^{\rm T}
    -\boldsymbol{{\rm{H}}}_k^s\boldsymbol{{\rm{X}}}_k^s{\boldsymbol{{\rm{Q}}}_k^s}^{\rm T}{\boldsymbol{{\rm{X}}}_k^s}^{\rm T}\\
    &\quad \quad \quad -\boldsymbol{{\rm{X}}}_k^s\boldsymbol{{\rm{Q}}}_k^s{\boldsymbol{{\rm{X}}}_k^s}^{\rm T}{\boldsymbol{{\rm{H}}}_k^s}^{\rm T})\\
    &- \theta_k^s tr(\boldsymbol{{\rm{X}}}_k^s {\boldsymbol{{\rm{X}}}_k^s}^{\rm T}
    -\boldsymbol{{\rm{X}}}_k^s{\boldsymbol{{\rm{C}}}_k^s}^{\rm T}{\boldsymbol{{\rm{T}}}^s}^{\rm T}
    -\boldsymbol{{\rm{T}}}^s\boldsymbol{{\rm{C}}}{\boldsymbol{{\rm{X}}}_k^s}^{\rm T})
  \end{aligned}
\end{equation}

Based on the conclusion of the above theorems, we can obtain:

\begin{equation}\label{eq_xglobalcommunitydecomposition_convergence_}
  \begin{aligned}
    &L(\boldsymbol{{\rm{X}}}_k^s) \\&
    \leq\alpha_k^s tr(-{\boldsymbol{{\rm{D}}}_k^s}^{\rm T}{{\boldsymbol{{\rm{X}}}_k^s}'}^{\rm T}\boldsymbol{{\rm{M}}}_k^s\boldsymbol{{\rm{Z}}}_k^s
    -{\boldsymbol{{\rm{D}}}_k^s}^{\rm T}{\boldsymbol{{\rm{Z}}}_k^s}^{\rm T}\boldsymbol{{\rm{M}}}_k^s{\boldsymbol{{\rm{X}}}_k^s}'\\
    &\quad \quad \quad  -{\boldsymbol{{\rm{D}}}_k^s}^{\rm T}{{\boldsymbol{{\rm{X}}}_k^s}'}^{\rm T}\boldsymbol{{\rm{M}}}_k^s{\boldsymbol{{\rm{X}}}_k^s}'-{\boldsymbol{{\rm{D}}}_k^s}^{\rm T}{{\boldsymbol{{\rm{X}}}_k^s}'}^{\rm T}{\boldsymbol{{\rm{M}}}_k^s}^{\rm T} \boldsymbol{{\rm{Z}}}_k^s\\
    &\quad \quad \quad  -\boldsymbol{{\rm{D}}}_k^s{{\boldsymbol{{\rm{X}}}_k^s}'}^{\rm T}{\boldsymbol{{\rm{M}}}_k^s}^{\rm T}{\boldsymbol{{\rm{X}}}_k^s}' \\
    &\quad \quad \quad +\frac{1}{2}(\boldsymbol{{\rm{R}}}_k^s  {\boldsymbol{{\rm{D}}}_k^s}^{\rm T}{{\boldsymbol{{\rm{X}}}_k^s}'}^{\rm T}{\boldsymbol{{\rm{X}}}_k^s}'\boldsymbol{{\rm{D}}}_k^s {{\boldsymbol{{\rm{X}}}_k^s}'}^{\rm T}\\
    &\quad \quad \quad +\boldsymbol{{\rm{R}}}_k^s  \boldsymbol{{\rm{D}}}_k^s{{\boldsymbol{{\rm{X}}}_k^s}'}^{\rm T}{\boldsymbol{{\rm{X}}}_k^s}'{\boldsymbol{{\rm{D}}}_k^s}^{\rm T} {{\boldsymbol{{\rm{X}}}_k^s}'}^{\rm T} ))\\
    &+\gamma_k^str(\frac{1}{2}(\boldsymbol{{\rm{R}}}_k^s  {{\boldsymbol{{\rm{X}}}_k^s}'}^{\rm T} {\boldsymbol{{\rm{H}}}^s}^{\rm T} {\boldsymbol{{\rm{H}}}^s}
    +{\boldsymbol{{\rm{H}}}^s}^{\rm T} {\boldsymbol{{\rm{H}}}^s}{\boldsymbol{{\rm{X}}}_k^s}'{{\boldsymbol{{\rm{X}}}_k^s}'}^{\rm T} )\\
    &\quad \quad \quad  -2{\boldsymbol{{\rm{Q}}}_{k,o}^s}^{\rm T} {\boldsymbol{{\rm{X}}}_{k,o}^s}^{\rm T} \boldsymbol{{\rm{H}}}^s\boldsymbol{{\rm{Z}}}_k^s\\
    &\quad \quad \quad  -2{\boldsymbol{{\rm{Q}}}_{k,o}^s}^{\rm T} {\boldsymbol{{\rm{X}}}_{k,o}^s}^{\rm T} \boldsymbol{{\rm{H}}}^s {{\boldsymbol{{\rm{X}}}_k^s}'}^{\rm T})\\
    &+\theta_k^s tr(\frac{1}{2}(\boldsymbol{{\rm{R}}}{{\boldsymbol{{\rm{X}}}_k^s}'}^{\rm T}+{\boldsymbol{{\rm{X}}}_k^s}'{{\boldsymbol{{\rm{X}}}_k^s}'}^{\rm T})\\
    &\quad \quad \quad  -2{\boldsymbol{{\rm{C}}}}^{\rm T} {\boldsymbol{{\rm{T}}}^s}^{\rm T}\boldsymbol{{\rm{Z}}}_k^s-2{\boldsymbol{{\rm{C}}}}^{\rm T} {\boldsymbol{{\rm{T}}}^s}^{\rm T}{\boldsymbol{{\rm{X}}}_k^s}')\\
    & =G(\boldsymbol{{\rm{X}}}_k^s,{\boldsymbol{{\rm{X}}}_k^s}')
  \end{aligned}
\end{equation}
where, $({\boldsymbol{{\rm{Z}}}_k^s})_{i,j}=({\boldsymbol{{\rm{X}}}_k^s}')_{i,j}\ln({\boldsymbol{{\rm{X}}}_k^s})_{i,j}/({\boldsymbol{{\rm{X}}}_k^s}')_{i,j}$, ${\boldsymbol{{\rm{R}}}}_{i,j}=({\boldsymbol{{\rm{X}}}_k^s})^4_{i,j}/({\boldsymbol{{\rm{X}}}_k^s}')^3_{i,j}$.

Then the parameter update rule Eq.\ref{eq_xosolver} of ${\boldsymbol{{\rm{X}}}_k^s}$ can be obtained by $\frac{\partial G({\boldsymbol{{\rm{X}}}_k^s},
{\boldsymbol{{\rm{X}}}_k^s}')}{\partial {\boldsymbol{{\rm{X}}}_k^s}}$. Because it meets the definition of the auxiliary function of Eq.\ref{eq_gassistant}. 
Therefore, the update rule can make the objective function of $\boldsymbol{{\rm{X}}}_k^s$ converge.

2) we obtain the Lagrangian function of $\boldsymbol{{\rm{D}}}_k^s$ according to Eq.\ref{eq_lagrangeglobalcommunitydecomposition}:

\begin{equation}\label{eq_dglobalcommunitydecomposition2_}
  \begin{aligned}
  &L(\boldsymbol{{\rm{D}}}_k^s)\\
  &=\alpha_k^s tr\Big(-\boldsymbol{{\rm{M}}}_k^s\boldsymbol{{\rm{X}}}_k^s{\boldsymbol{{\rm{D}}}_k^s}^{\rm T}{\boldsymbol{{\rm{X}}}_k^s}^{\rm T}
  -\boldsymbol{{\rm{X}}}_k^s\boldsymbol{{\rm{D}}}_k^s{\boldsymbol{{\rm{X}}}_k^s}^{\rm T}{\boldsymbol{{\rm{M}}}_k^s}^{\rm T}\\
  &\quad \quad \quad +\boldsymbol{{\rm{X}}}_k^s\boldsymbol{{\rm{D}}}_k^s{\boldsymbol{{\rm{X}}}_k^s}^{\rm T}\boldsymbol{{\rm{X}}}_k^s{\boldsymbol{{\rm{D}}}_k^s}^{\rm T}{\boldsymbol{{\rm{X}}}_k^s}^{\rm T}\Big)\\
  &+\gamma_k^s \Big(\sum_j\sum_n((\boldsymbol{{\rm{H}}}^s\boldsymbol{{\rm{X}}}_k^s)_{j,n})^2 \big( \underbrace{ \sum_i\sum_m(\boldsymbol{{\rm{X}}}_k^s)_{i,m}({\boldsymbol{{\rm{D}}}_k^s}')_{m,n} \frac{({\boldsymbol{{\rm{D}}}_k^s})_{m,n}}{({\boldsymbol{{\rm{D}}}_k^s}')_{m,n}} }_{(\boldsymbol{{\rm{Q}}}_k^s)_{n,n}} \big)^2\\
  &\quad \quad-2\sum_j\sum_n(\boldsymbol{{\rm{H}}}^s\boldsymbol{{\rm{X}}}_k^s)_{j,n}  \underbrace{ \sum_i\sum_m(\boldsymbol{{\rm{X}}}_k^s)_{i,m}({\boldsymbol{{\rm{D}}}_k^s}')_{m,n}\frac{({\boldsymbol{{\rm{D}}}_k^s})_{m,n}}{({\boldsymbol{{\rm{D}}}_k^s}')_{m,n}} }_{(\boldsymbol{{\rm{Q}}}_k^s)_{n,n}} \\
  & \quad \quad \quad (\boldsymbol{{\rm{Q}}}_{k,o}^s)_{n,n}(\boldsymbol{{\rm{X}}}_{k,o}^s)_{j,n}\Big)\\
  &+\theta^s_k \Big(\sum_j\sum_n(\boldsymbol{{\rm{X}}}_k^s)^2_{j,n}\big( \underbrace{  \sum_i\sum_m(\boldsymbol{{\rm{X}}}_k^s)_{i,m}({\boldsymbol{{\rm{D}}}_k^s}')_{m,n}\frac{({\boldsymbol{{\rm{D}}}_k^s})_{m,n}}{({\boldsymbol{{\rm{D}}}_k^s}')_{m,n}} }_{(\boldsymbol{{\rm{Q}}}_k^s)_{n,n}} \big)^2\\
  & -2\sum_j\sum_n(\boldsymbol{{\rm{X}}}_k^s)_{j,n} \underbrace{  \sum_i\sum_m(\boldsymbol{{\rm{X}}}_k^s)_{i,m}({\boldsymbol{{\rm{D}}}_k^s}')_{m,n}\frac{({\boldsymbol{{\rm{D}}}_k^s})_{m,n}}{({\boldsymbol{{\rm{D}}}_k^s}')_{m,n}}} _{(\boldsymbol{{\rm{Q}}}_k^s)_{n,n}}(\boldsymbol{{\rm{T}}}^s\boldsymbol{{\rm{C}}})_{j,n}\Big)
  \end{aligned}
\end{equation} 

Based on the conclusion of the theorem, we can get:

\begin{equation}\label{eq_dglobalcommunitydecomposition_convergence_}
  \begin{aligned}
    &L(\boldsymbol{{\rm{D}}}_k^s)\\ 
    &\leq \alpha_k^s tr\Big(-2{\boldsymbol{{\rm{X}}}_k^s}^{\rm T}{\boldsymbol{{\rm{M}}}_k^s}^{\rm T}\boldsymbol{{\rm{X}}}_k^s\boldsymbol{{\rm{Z}}}_k^s
     -2{\boldsymbol{{\rm{X}}}_k^s}^{\rm T}{\boldsymbol{{\rm{M}}}_k^s}^{\rm T}\boldsymbol{{\rm{X}}}_k^s{\boldsymbol{{\rm{D}}}_k^s}'\\
    &\quad \quad  +\frac{1}{2}( {\boldsymbol{{\rm{X}}}_k^s}^{\rm T} \boldsymbol{{\rm{X}}}_k^s  {\boldsymbol{{\rm{Y}}}_k^s}^{\rm T} {\boldsymbol{{\rm{X}}}_k^s}^{\rm T} \boldsymbol{{\rm{X}}}_k^s   {\boldsymbol{{\rm{D}}}_k^s}'\\
    &\quad \quad  + {\boldsymbol{{\rm{X}}}_k^s}^{\rm T} \boldsymbol{{\rm{X}}}_k^s  {{\boldsymbol{{\rm{D}}}_k^s}'}^{\rm T} {\boldsymbol{{\rm{X}}}_k^s}^{\rm T} \boldsymbol{{\rm{X}}}_k^s \boldsymbol{{\rm{Y}}}_k^s )\Big)\\
    &+\gamma_k^s  \Big(\sum_j\sum_n((\boldsymbol{{\rm{H}}}^s\boldsymbol{{\rm{X}}}_k^s)_{j,n})^2   \underbrace{\sum_i\sum_m(\boldsymbol{{\rm{X}}}_k^s)_{i,m}({\boldsymbol{{\rm{D}}}_k^s}')_{m,n}}_{(\boldsymbol{{\rm{Q}}}_k^s)_{n,n}} \\
    &\quad \quad  \sum_i\sum_m\Big( (\boldsymbol{{\rm{X}}}_k^s)_{i,m}({\boldsymbol{{\rm{D}}}_k^s}')_{m,n}(\frac{({\boldsymbol{{\rm{D}}}_k^s})_{m,n}}{({\boldsymbol{{\rm{D}}}_k^s}')_{m,n}})^2 \Big)\\
    &\quad \quad  -2\sum_j\sum_n(\boldsymbol{{\rm{H}}}^s\boldsymbol{{\rm{X}}}_k^s)_{j,n}\sum_i\sum_m(\boldsymbol{{\rm{X}}}_k^s)_{i,m}({\boldsymbol{{\rm{D}}}_k^s}')_{m,n}\\
    &\quad \quad  (1+\ln \frac{({\boldsymbol{{\rm{D}}}_k^s})_{m,n}}{({\boldsymbol{{\rm{D}}}_k^s}')_{m,n}})(\boldsymbol{{\rm{Q}}}_{k,o}^s)_{n,n}(\boldsymbol{{\rm{X}}}_{k,o}^s)_{j,n}\Big)\\
    &+\theta^s_k  \Big(\sum_j\sum_n(\boldsymbol{{\rm{X}}}_k^s)^2_{j,n}  \underbrace{\sum_i\sum_m(\boldsymbol{{\rm{X}}}_k^s)_{i,m}({\boldsymbol{{\rm{D}}}_k^s}')_{m,n}}_{(\boldsymbol{{\rm{Q}}}_k^s)_{n,n}} \\
    &\quad \quad  \sum_i\sum_m \Big((\boldsymbol{{\rm{X}}}_k^s)_{i,m}({\boldsymbol{{\rm{D}}}_k^s}')_{m,n} (\frac{({\boldsymbol{{\rm{D}}}_k^s})_{m,n}}{({\boldsymbol{{\rm{D}}}_k^s}')_{m,n}})^2\Big)\\
    &\quad \quad  -2\sum_j\sum_n(\boldsymbol{{\rm{X}}}_k^s)_{j,n}\sum_i\sum_m(\boldsymbol{{\rm{X}}}_k^s)_{i,m}({\boldsymbol{{\rm{D}}}_k^s}')_{m,n}\\
    &\quad \quad  (1+\ln\frac{({\boldsymbol{{\rm{D}}}_k^s})_{m,n}}{({\boldsymbol{{\rm{D}}}_k^s}')_{m,n}})(\boldsymbol{{\rm{T}}}^s\boldsymbol{{\rm{C}}})_{j,n}\Big)\\
    & =G(\boldsymbol{{\rm{D}}}_k^s,{\boldsymbol{{\rm{D}}}_k^s}')
  \end{aligned}
\end{equation}

Then, the parameter update rule Eq.\ref{eq_dsolver} of $\boldsymbol{{\rm{D}}}_k^s$ can be obtained by $\frac{\partial G({\boldsymbol{{\rm{D}}}_k^s},
{\boldsymbol{{\rm{D}}}_k^s}')}{\partial {\boldsymbol{{\rm{D}}}_k^s}}$. It also meets the auxiliary function definition of Eq.\ref{eq_gassistant}. 
Therefore, the update rule can make equation (8) converge.


 





\end{document}